\documentclass[11pt]{article}
\usepackage[top=2.5cm,bottom=2.5cm,left=2.55cm,right=2.55cm]{geometry}
\usepackage{rotating}

\usepackage{enumitem}

\usepackage{bbm}
\usepackage{overpic}
\usepackage{caption}
\usepackage{mathtools}
\usepackage{latexsym} 
\usepackage{verbatim}  
\usepackage{tikz}
\usetikzlibrary{matrix}
\usepackage{color}
\usepackage{graphicx,amssymb,amsfonts,amsmath,amssymb,amscd,amstext, mathrsfs}
\usepackage{graphicx}
\usepackage{bbm}
\usepackage{dsfont}
\usepackage{amsmath}
\usepackage{empheq}
\usepackage{cite}
\usepackage{tikz}
\usetikzlibrary{shapes.misc, positioning,decorations.pathreplacing,angles,quotes}
\usetikzlibrary{decorations.pathmorphing}
\usepackage{soul} 
\definecolor{blul}{RGB}{51, 153,255}
\definecolor{greenl}{RGB}{76,153,0}
\definecolor{violetto}{RGB}{0, 13, 122}

\numberwithin{equation}{section}

\newlength\dlf

\newcommand{\bw}{\begin{widetext}}
\newcommand{\ew}{\end{widetext}}
\newcommand{\bea}{\begin{eqnarray}}
\newcommand{\eea}{\end{eqnarray}}

\renewcommand{\bar}[1]{\overline{#1}}
\renewcommand{\tilde}[1]{\widetilde{#1}}
\renewcommand{\hat}[1]{\widehat{#1}}

\newcommand{\<}{\langle}
\renewcommand{\>}{\rangle}

\renewcommand{\cal}{\mathcal}

\newcommand{\Dmax}{\Delta_{\max}}

 \definecolor{themecolor}{RGB}{0, 64, 168}
\definecolor{myRed}{RGB}{150,22,22}
\definecolor{myRedL}{RGB}{255,239,237}
\definecolor{myGrayL}{RGB}{220,220,220}
\definecolor{blul}{RGB}{51, 153,255}
\definecolor{greenl}{RGB}{76,153,0}

\DeclareFontShape{OT1}{cmr}{mx}{n}{<->cmr10}{}
\newcommand{\titlefont}{\fontseries{mx}\selectfont}

\def\frac#1#2{{#1\over #2}}

\usepackage{subcaption}
\usepackage{graphicx}
\usepackage{mymacroMOD}
\usepackage{jheppub}

\usepackage{braket}
\usetikzlibrary{decorations.pathmorphing}
\tikzset{snake it/.style={decorate, decoration=snake}}
\usetikzlibrary{arrows.meta}
\makeatletter
\newlength\lrvec@height
\newlength\lrvec@width
\newif\iflrvec@same@height
\def\lrvec{\@ifstar\slrvec@\lrvec@}
\newcommand{\slrvec@}[2][.4ex]{
  \lrvec@same@heighttrue
  \mathpalette\lrvec@@{{#1}{#2}}
}
\newcommand{\lrvec@}[2][.4ex]{
  \lrvec@same@heightfalse
  \mathpalette\lrvec@@{{#1}{#2}}
}
\def\lrvec@@#1#2{\lrvec@@@#1#2}
\def\lrvec@@@#1#2#3{%
  \iflrvec@same@height
    \settoheight{\lrvec@height}{$\m@th#1 \mathbf{T}#3$}
  \else
    \settoheight{\lrvec@height}{$\m@th#1#3$}
  \fi
  \settowidth{\lrvec@width}{$\m@th#1#3$}
  \kern.08em
  \raisebox{#2}{\raisebox{\lrvec@height}{\rlap{%
    \kern-.05em
    \begin{tikzpicture}[<-> /.tip={To[width=.4em, length=.2em]}]
      \draw [<->] (-.05em,0)--(\lrvec@width+.05em,0);
    \end{tikzpicture}%
  }}}%
  #3
  \kern.08em
}
\makeatother

\usepackage{simpler-wick}
\graphicspath{ {./figure/} }

\newcommand\blfootnote[1]{%
  \begingroup
  \renewcommand\thefootnote{}\footnote{#1}%
  \addtocounter{footnote}{-1}%
  \endgroup
}

\usepackage{subcaption}

\begin{document}
\pagestyle{myplain}
\begin{titlepage}

\begin{flushright} 
\end{flushright}

\begin{center} 

\vspace{0.35cm}

{\fontsize{20.5pt}{25pt}
{\titlefont  
3d Ising Field Theory with Magnetic Deformation: Fuzzy Sphere Meets TCSA
}}

\vspace{1.6cm}  

{{Giulia Fardelli$^a$\blfootnote{${}^a$\href{mailto:fardelli@bu.edu}{\tt fardelli@bu.edu}}, A. Liam Fitzpatrick$^a$\blfootnote{${}^a$\href{mailto:fitzpatr@bu.edu}{\tt fitzpatr@bu.edu}},  Emanuel Katz$^{a}$\blfootnote{${}^{a}$\href{mailto:amikatz@bu.edu}{\tt amikatz@bu.edu}},  Yuan Xin$^{b,c}$\blfootnote{${}^{c}$\href{mailto:yuanxin@andrew.cmu.edu}{\tt xinyuan@simis.cn}}
}

\vspace{1cm} 

{{\it
${}^a$Department of Physics, Boston University, 
Boston, MA  02215, USA
}}\\

{\it
${}^b$Center for Mathematics and Interdisciplinary Sciences, Fudan University, Shanghai
200433, China
}\\

{\it
${}^c$Shanghai Institute for Mathematics and Interdisciplinary Sciences (SIMIS), Shanghai
200433, China
}}\\
\end{center}
\vspace{1.5cm}

{\noindent 
We study the magnetic deformation of the $(2+1)$d Ising CFT, also known as Ising Field Theory (IFT), on a spatial sphere.  We use the Fuzzy Sphere (FS) Ising setup and compare it to the Truncated Conformal Space Approach (TCSA).  
Universal, infinite volume, IFT quantities are extracted by modeling the effects of curvature, as expected from a local effective theory on the sphere.  We find evidence that IFT contains a bound-state with a small binding energy.  Locality also allows us to extract the same IFT quantities from different angular momentum sectors, providing additional consistency checks.}

\end{titlepage}

\tableofcontents

\newpage

\newpage

\section{Introduction and Summary} 

A modern picture of Quantum Field Theories (QFTs) is that they are points along Renormalization Group (RG) flows.  The endpoints in the infrared (IR) can be gapless Conformal Field Theories (CFTs) or theories with a mass gap, whereas the ultraviolet (UV) can be a wide variety of microscopic systems.  For instance,  the theory at the shortest distances can be a lattice model or a non-relativistic collection of particles.  The 2d and 3d Ising CFTs hold a special place in this landscape of networks of RG flows, as the unitary CFTs with the smallest possible nonzero central charge in their respective dimension \cite{El-Showk:2014dwa}. 
As a result, they tend to be the simplest CFTs to find when tuning a UV theory to a critical point where the gap vanishes.  
Moreover, once the Ising CFT is found by such a tuning, de-tuning slightly away from the fixed point still gives a nontrivial RG flow that passes first through Ising CFT\footnote{More accurately, the RG flow passes arbitrarily close to the Ising CFT.} on its way into the deeper IR where the mass gap from the detuning becomes visible.  This vicinity of the Ising CFT was named `Ising Field Theory' by Zamolodchikov~\cite{Zamolodchikov:1989hfa}, who studied it in detail in 2d.  

From a field theorist's perspective, what is most appealing about Ising Field Theory is that, like the Ising CFT itself, up to a single dimensionless ratio, it depends only on macroscopic properties of the model and not on any of the details of the UV theory that was tuned to reach the critical point. In fact, Ising Field Theory does not require any additional dynamical data to define the theory beyond the CFT data -- OPE coefficients and scaling dimensions of operators -- that characterize the Ising CFT itself.  The reason for this is that one can define the gapped vicinity of the critical point by the Ising CFT plus its two relevant deformations $\sigma$ and $\epsilon$:
\begin{equation}
H_{\rm IFT} = H_{\rm Ising-CFT} +\frac{1}{V_d} \int d^{d-1} x \left( g \lsp \epsilon(x) + h_z \lsp \sigma(x)\right),
\label{eq:IFT-Ham}
\end{equation}
where $V_d = \frac{2\pi^{d/2}}{\Gamma(d/2)}$ is the volume of $S^{d-1}$. 
The local operators $\epsilon$ and $\sigma$ are defined as the two lowest-dimension nontrivial (i.e. not the identity) operators that are even and odd, respectively, under the $\mathbb{Z}_2$ symmetry of the theory.   This description flips the perspective of the Ising CFT as the IR fixed point and places it as the UV fixed point, with Ising Field Theory as the RG flow emanating from it.  All dimensionless ratios in IFT are a function only of the dimensionless ratio $g^{\frac{1}{d-\Delta_\epsilon}}/h_z^{\frac{1}{d-\Delta_\sigma}}$ of the two `couplings' $g$ and $h_z$. What is often less-appreciated is the fact that, given only the value of these couplings and the CFT data of the Ising CFT, it is possible in practice to calculate the energy spectrum of the gapped theory nonperturbatively, using the Truncated Conformal Space Approach (TCSA)~\cite{Yurov:1989yu,Hogervorst:2014rta}, as we will review.   This general construction, of QFTs defined by CFTs plus their relevant deformations, has many virtues as a definition of QFTs quite broadly.  One advantage is that in this construction the QFT is not emergent from some underlying medium or UV system, but rather is formulated entirely in terms of intrinsic QFT degrees of freedom, in this case the Hilbert space of the CFT.  
 It can also potentially be more efficient as it is working directly with IR states. However until now all applications of TCSA have been to deformations of integrable CFTs.  One of our main motivations for studying IFT in 3d in particular, is that it is probably one of the simplest examples where we can study the performance of TCSA with a non-integrable CFT UV, while at the same time comparing our results directly to another non-perturbative Hamiltonian method, namely the Fuzzy Sphere~\cite{He:2026ong}.

In terms of the framework above, the Fuzzy Sphere (FS) is a UV non-relativistic model of $N$ interacting fermions that at criticality flows to the 3d Ising CFT in the IR~\cite{Zhu:2022gjc}.  For our purposes, it can be thought of as an ideal regulator to study both IFT and TCSA simultaneously.  What makes the FS special as a microscopic model is that it exactly preserves the rotational symmetry of the theory on the spatial sphere. 
Thus, at criticality, its IR states become approximations of CFT states, labelled by the exact same quantum numbers as they would be in the Ising CFT.  Moreover, it is straightforward to find microscopic FS local operators which flow in the IR to desired scalar primary operators, while preserving rotational invariance.  Engineering~(\ref{eq:IFT-Ham}) then simply involves slight detuning from criticality as well as an addition of convenient $\mathbb{Z}_2$-odd operator to the FS Hamiltonian.  More concretely, IFT is described in the FS framework by a Hamiltonian of the form
\begin{equation}
H_{\rm FS} = H_{\rm fp} + \int d^2\Omega \left( (h_x-h_x^c) \psi^\dagger \sigma^x \psi + h_{z, \text{FS}} \psi^\dagger \sigma^z \psi \right),
\end{equation}
where $\int d^2\Omega$ is the integral over the two-sphere, $\psi$ is a non-relativistic spin-$1/2$ fermion in the presence of a magnetic flux through the sphere, and $\sigma^x, \sigma^z$ are Pauli matrices, while $h_x^c$ is the value of the transverse field at criticality.  At low energies, the operator $\psi^\dagger \sigma^z \psi $ is dominantly $\sigma$ and $\psi^\dagger \sigma^x \psi $ is dominantly $\epsilon$ (plus a constant), so that this Hamiltonian describes the same low-energy parameter space of theories as (\ref{eq:IFT-Ham}).

In this paper, we will restrict our attention to IFT with only the $\sigma$ deformation, so $g=0$ in~\eqref{eq:IFT-Ham}, and take space to be a sphere with radius $R$.  In this case, all dependence of the energy spectrum on $h_z$ is tightly constrained by dimensional analysis.  In particular, we can parameterize the vacuum energy $E_{\rm vac}$ and the excited states $E_i = E_{\rm vac} + m_i$ in the spin-zero sector as
\begin{equation}
E_{\rm vac} = 4 \pi R^2 {\cal E}_{\rm vac}  = \mu^3 R^2 f_0(\mu  R), \qquad m_i  = \mu f_i(\mu R), \quad \text{with}\quad 
\mu  \equiv  h_z^{\frac{1}{3-\Delta_\sigma}} .
\label{eq:FiniteVolumeDimAnalysis}
\end{equation}
Intensivity of the vacuum energy density ${\cal E}_{\rm vac}$ and mass gaps $m_i$ implies that at large coupling $h_z$, the functions $f_n$ defined this way approach constants:
\begin{equation}
\lim_{h_z \rightarrow \infty} f_n(\mu R) = A_n,
\label{eq:AiDef}
\end{equation}
and one of our primary goals will be to calculate them numerically.  On general grounds, the mass $m_1$ of the lightest one-particle state implies that at large volume, there is a two-particle continuum starting at $2m_1$.  Any excited states in the range $m_1 < m_i < 2m_1$ are bound states that are stable kinematically.

Using the FS to directly study IFT, we must be mindful of the fact that all observables receive corrections due to irrelevant operators:
\begin{equation}
\label{eq:HFS-EFT}
H_{\rm FS}  = H_{\rm IFT} + \int d^2\Omega \sum_{\rm primaries} \frac{g_\CO}{(\Lambda_{\rm UV})^{\Delta_\CO-3}} \CO(\Omega)\, ,
\end{equation}
with $\Delta_{\CO}$ the conformal dimension of the $\CO$ operator and $g_{\CO}$ Wilson coefficients.
Consequently, we expect corrections from the leading irrelevant operators which scale as $\left(\frac{\mu}{\Lambda_{\rm UV}}\right)^{\Delta_\CO-3}$, with $\Lambda_{\rm UV} \sim \sqrt{N}$.  Hence, when $h_z$ becomes too large, the above parameterization will break down.  On the other hand, in the regime where the corrections can be dominantly attributed to a few leading operators (in our case, the spin four operator $C_{\mu_1\cdots \mu_4}$ and the spin two operator $T^\prime_{\mu_1\mu_2}$, both of dimension $\sim 5$) we can reliably extrapolate the observables to $N=\infty$.\footnote{We must also keep track of curvature corrections, coming from operators which include the Ricci scalar, $\CR^n \CO$, such that $g_\CO$ is itself an expansion in powers of $1/N$, see the discussion in~\cite{Fardelli:2026zas,Lauchli:2025fii} for more details.}

We can also study IFT using TCSA techniques, using Ising CFT data which we recently extracted using the FS~\cite{Fardelli:2026zas}.\footnote{See also \cite{Caselle:1999tm,Caselle:2001im} for studies of 3d IFT with explicit $\mathbb{Z}_2$-breaking using a lattice formulation.}  In this approach, we compute the Hamiltonian in the CFT basis of states and truncate to states with dimensions/energies below a certain $\Delta_{\rm max}$.  Ideally, we would also extrapolate our observables in $\Delta_{\rm max}$.  However, since the truncation yields non-local corrections, a proper extrapolation requires better knowledge of Ising CFT correlators.  Thus, for now, we will be content with raw un-extrapolated TCSA data, which nevertheless we will compare with direct FS IFT results.  FS IFT results also allow us to perform a more indirect check of TCSA:  Namely, we can use the FS to decompose IFT states in terms of the CFT basis states at the critical point.  In other words, we can measure the content of particular state at a given $h_z$ in terms of states at $h_z=0$.  

Before launching into a detailed analysis, let us summarize a few of our quantitative results here.  First, after canonically normalizing the $\sigma(x)$ operator as $\langle \sigma(x) \sigma(0)\rangle = 1/x^{2\Delta_\sigma}$, we find that the mass-gap (or inverse correlation length) is related to the $\sigma$ one point function as:
\begin{equation}
\langle \sigma \rangle  =  \partial_{h_z} E_0  \approx   - 1.28 \   m_{\rm gap}^{\Delta_{\sigma}},
\end{equation}
where $E_0 \approx -1.37 h_z^{\frac{3}{3-\Delta_\sigma}}$ is the sphere vacuum energy, and we find the gap is related to the coupling $h_z$ by   $m_{\rm gap} \approx 1.65 h_z^{\frac{1}{3-\Delta_\sigma}}$.   
We also find a bound state in the spectrum, with a small binding energy.  Our best-fit value of the bound state mass in units of the mass gap is $m_{\rm BS}/m_{\rm gap} \approx 1.96$.\\

Our paper is organized as follows: We begin with a review of IFT in Sec.~\ref{sec:IFTreview}, including analytic results in 2d and 3d.  The FS setup is then briefly described in Sec.~\ref{sec:FSsetup}.  In Sec.~\ref{sec:Hierarchy} we discuss the local EFT, and its various scales, that we use to fit our numerical data.  In Sec.~\ref{sec:FSresults} we report on our results using the FS formulation of the IFT.  We then use the CFT data directly using the TCSA formulation of IFT and compare our results to the previous sections in Sec.~\ref{sec:TCSAresults}.  Finally, we discuss some future directions for investigation in Sec.~\ref{sec:future}.  The appendices contain further technical details and collect the data used in our analysis.\\

\textbf{Note added}: Toward the completion of this paper,~\cite{Taylor:2026wan} appeared, with some overlap with the results of Sec.~\ref{sec:FSresults}. We briefly comment on the relation between the two works at the end of that section.

  \begin{figure}\centering
\begin{minipage}{0.48\textwidth}\centering
 \includegraphics[width=1\textwidth]{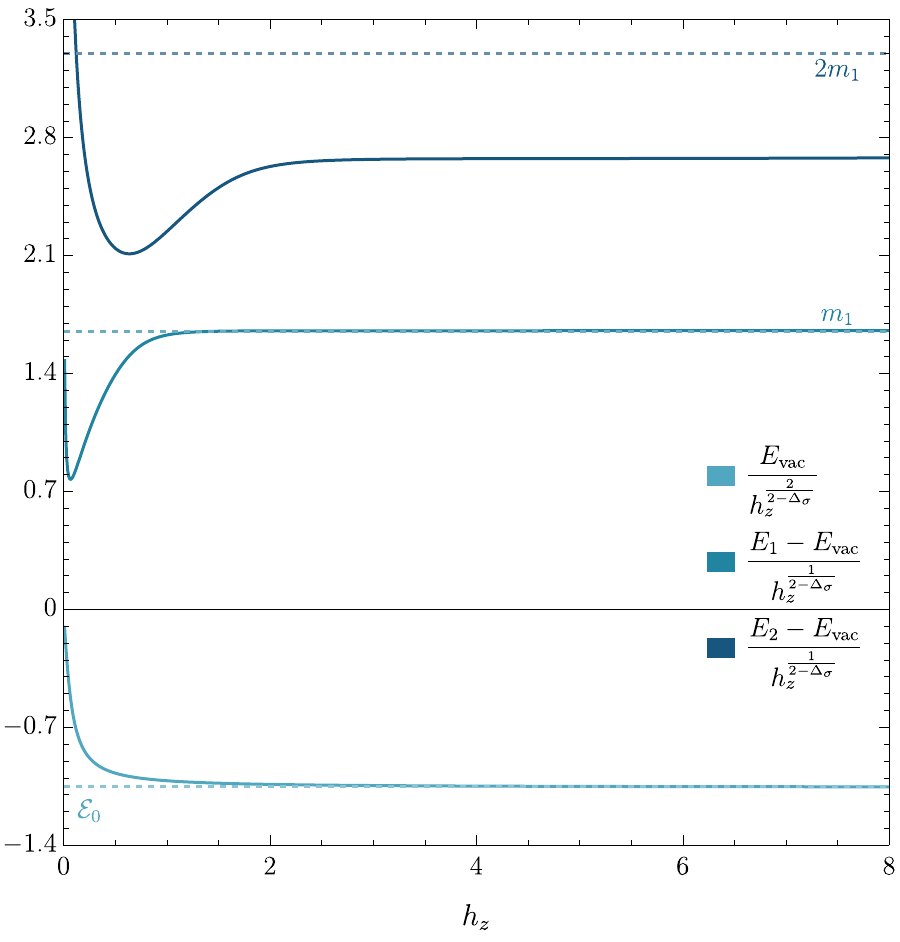}
 \end{minipage}
 \hfill
 \begin{minipage}{0.48\textwidth}\centering
 \includegraphics[width=1\textwidth]{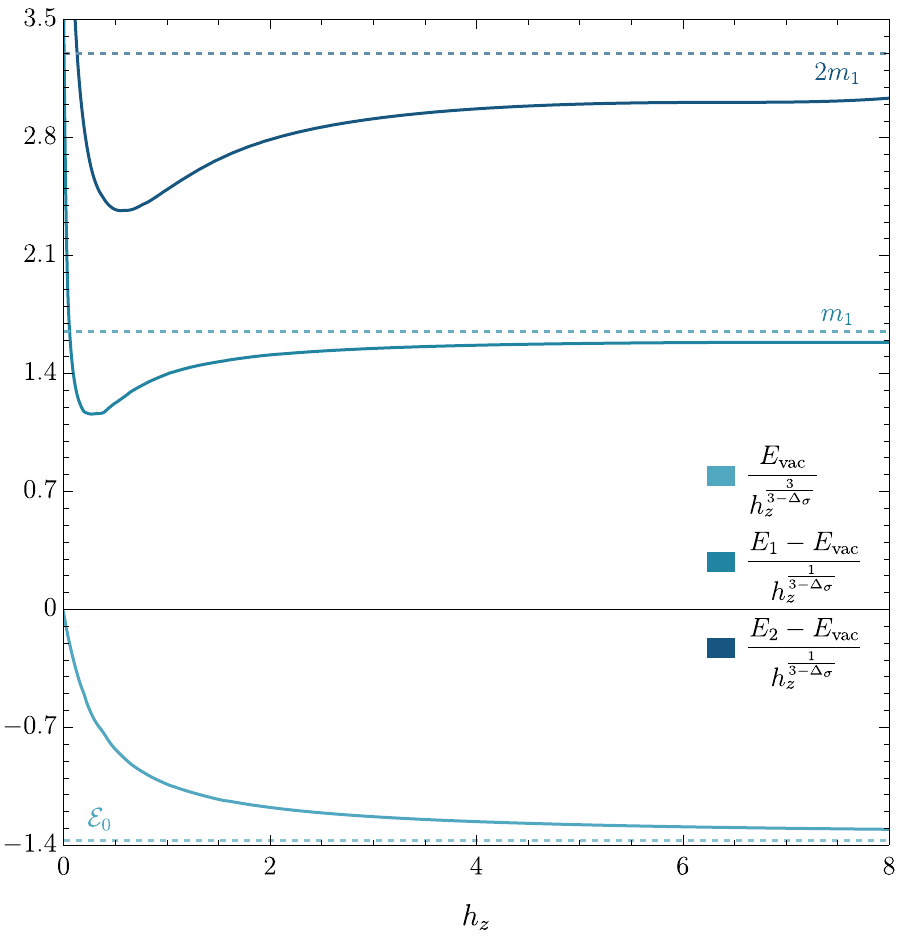}
 \end{minipage}
 \caption{Spectrum of Ising Field theory in 2d (\textit{left}) and in 3d (\textit{right}) including the vacuum energy,  mass gap and bound state divided by the appropriate power of mass $\mu=h_z^{\frac{1}{d-\Delta_{\sigma}}}$. 
 In 2d, the finite volume corrections to the energy differences $E_i - E_{\rm vac}$ are exponentially suppressed at large $\mu$ whereas in 3d the corrections are power-law suppressed.
 }\label{fig:2dVs3dTCSA}
 \end{figure} 
\section{Ising Field Theory Review}
\label{sec:IFTreview}
\subsection{Brief Review of 2d Ising Field Theory}

We begin with a brief review of 2d Ising Field Theory with the magnetic $\sigma$ deformation only.  The lattice description of the usual Ising model is:
\begin{equation}
H = -J \sum_{i=1}^N \left( \sigma^z_i \sigma^z_{i+1} +g\lsp \sigma^x_i + h  \lsp \sigma^z_i\right),
\end{equation}
where $g=1$, $h=0$ is the critical point and $i=N+1$ is the same site as $i=1$.  Because a periodic lattice with $N$ sites preserves a $\mathbb{Z}_N$ subgroup of the $SO(2)$ rotation group, the fuzzy sphere technology is not a significant advantage in 2d.  It will be useful to review some standard results from the 2d case to set the stage for analogous 3d results in the remainder of the paper.

 At the critical point, in the large $N$ limit the IR of the theory flows to the 2d Ising CFT.  Setting $J=\frac{N}{4\pi}$ rescales the energies to match the operator spectrum of the 2d Ising CFT.   To study Ising Field Theory with the magnetic deformation, in the CFT description one deforms by adding the CFT primary operator $\sigma$ to the Hamiltonian,
 \begin{equation}
H = H_{\rm 2d  Ising} + \frac{h_z}{2\pi} \int_0^{2\pi}  d x \,  \sigma(x),
\end{equation}
where $\sigma$ is a CFT primary operator normalized so that at $h_z=0$, $\langle \sigma(x) \sigma(y)\rangle \stackrel{x \sim y}{\sim} |x-y|^{-1/4}$.  
 In the lattice description the simplest operator that flows to $\sigma$ in the IR is the lattice operator $\sigma_z^i$.  At large separation $|i-j| \gg 1$ on an infinite length chain,
\begin{equation}
\< \sigma_i^z \sigma_j^z \> \sim \frac{a_\sigma^2}{|i-j|^{1/4}} ,
\end{equation}
The coefficient $a_\sigma = 2^{1/24}e^{1/8} A_G^{-3/2} = 0.80312044$ is known exactly \cite{wu1966theory}.\footnote{$A_G \approx 1.28243$ is the Glaisher-Kinkelin constant.} The value of $a_\sigma$ is only necessary for converting the coefficient $h$ in the lattice description to the coupling $h_z$ in the field theory description of the Hamiltonian.

Many remarkable closed-form results are known about the field theory near the critical point with only the magnetic $\sigma$ deformation.  This limit is sometimes referred to as the ``$E_8$'' theory due to the spectrum of bound states.  For comparison with our 3d results, the relevant 2d results are the vacuum energy and the first few one-particle state masses~\cite{fateev1994exact}:
\begin{equation}
\begin{aligned}
E_{\rm vac} &=  A_0 h_z^{\frac{2}{2-\Delta_\sigma}} , \qquad A_0 = -\frac{\Gamma \mleft(\frac{1}{5}\mright) \Gamma \mleft(\frac{1}{3}\mright) \Gamma
   \mleft(\frac{7}{15}\mright)}{\Gamma \mleft(\frac{8}{15}\mright) \Gamma
   \mleft(\frac{2}{3}\mright) \Gamma \mleft(\frac{4}{5}\mright)} \left( \frac{\Gamma \mleft(\frac{3}{4}\mright) \Gamma \mleft(\frac{13}{16}\mright)^2}{\Gamma
   \mleft(\frac{3}{16}\mright)^2 \Gamma \mleft(\frac{1}{4}\mright)}\right)^{\frac{8}{15}} \approx  -1.05962   , \\
m_1 &= A_1 h_z^{\frac{1}{2-\Delta_\sigma}}, \qquad A_1 =4  \sqrt{\frac{|A_0|}{\pi}   \sin\left(\frac{\pi}{3}\right) \sin\left(\frac{\pi}{5}\right) \cos\left(\frac{\pi}{30}\right)} \approx 1.65288 , \\
m_2/m_1 &= 2 \cos \left(\frac{\pi}{5}\right) \approx 1.61803 .
\end{aligned}
\end{equation}
Each one of these quantities has an exactly analogous version in 3d, and the quantities $A_0$ and $A_1$ are just like the 3d $A_n$ values from~(\ref{eq:AiDef}).  Because the flat-space limit of IFT has only a single dimensionful scale, set by the size of the coefficient $h_z$ of the relevant deformation, by dimensional analysis this coupling factors out of all dimensionless quantities like the values $A_0, A_1$ and $m_2/m_1$.  In other words, like pure QCD, it is a theory without any free parameters.  It is a strongly coupled theory in the infrared and all three of these quantities must be computed using nonperturbative methods.

This model was  one of the first examples that Zamolodchikov and Yurov~\cite{Yurov:1991my} studied with the TCSA method.  Because the 2d Ising CFT is solvable, it is possible to efficiently compute the matrix elements of $\sigma$ (as well as the $\mathbb{Z}_2$-even relevant deformation $\epsilon$) between any bra and ket state in the CFT limit in radial quantization,\footnote{In fact one can also compute the matrix elements between any bra and ket state in the QFT with the massive $\epsilon$ deformation turned on as well \cite{Fonseca:2006au,Zamolodchikov:2013ama, Gabai:2019ryw}. } and use this to numerically determine the observables in the theory.  The availability of efficient lattice and field theory methods for numerically studying the theory, as well as the existence of the integrable results, make the model  an essentially perfect playground for exploring nonperturbative techniques.  At \href{https://github.com/andrewliamfitz/2dIFT-TCSA}{https://github.com/andrewliamfitz/2dIFT-TCSA}, we provide a notebook implementing TCSA for the 2d Ising model for both $\sigma$ and $\epsilon$ deformations, which can be used to reproduce the results in Fig.~\ref{fig:2dVs3dTCSA}.

\subsection{3d Ising Field Theory Perturbative Results}

Much less is known about the spectrum of the magnetic deformation of  3d Ising Field Theory.  On a two-sphere of radius $R$, at small values of the coupling $h_z$ one can compute the spectrum order-by-order in $h_z$ as integrals over the correlation functions of $\sigma$ in flat space $\mathbb{R}^3$.  For instance, the vacuum energy at all orders in $h_z$ is
\begin{equation}
\begin{aligned}
E_{\rm vac}(h_z) &=- \sum_{n=2}^\infty \left( \frac{h_z}{V_d} \right)^{n} C_n , \\
C_n &= \frac{(-1)^n}{n!} \Big(V_d \prod_{i=1}^{n-1}  \int \frac{d^d x_i}{ x_i^y} \Big) \Big\langle \sigma(x_1) \dots \sigma(x_n) \Big\rangle_{\rm conn}
\end{aligned}
\end{equation}
where $y=d-\Delta_\sigma$, $V_d = \frac{2 \pi^{\frac{d}{2}}}{\Gamma(\frac{d}{2})}$ is the volume of the $(d-1)$-sphere, the correlator and integrals are on $\mathbb{R}^3$ with $x_n$ fixed to a reference point on the unit sphere, and we have written the formula for general $d$.  

At $O(h_z^2)$, the vacuum energy just depends on an integral of the two-point function, which can be done in closed form:
\begin{equation}
E_{\rm vac}(h_z) \supset  -h_z^2 \frac{C_2}{V_3^2} =
 - h_z^2  \frac{\sqrt{\pi} \Gamma\mleft(\frac{3}{2}-\Delta_{\sigma}\mright)\Gamma\mleft(\frac{\Delta_{\sigma}}{2}\mright)^2}{8 \Gamma\mleft(\frac{3-\Delta_{\sigma}}{2}\mright)^2\Gamma\mleft(\Delta_{\sigma}\mright)}= -1.933 h_z^2\, .
   \label{eq:CPTQuadVac}
   \end{equation}
Going to higher orders is more challenging since it requires higher order correlators.  At $O(h_z^4)$, one can use the known OPE coefficients in the 3d Ising CFT to get a fairly accurate calculation of the $\sigma$ correlator.  Including the conformal blocks for $\epsilon, \epsilon^\prime, \epsilon^{\prime\prime}, T, T^\prime$ and $C$ we find
\begin{equation}
E_{\rm vac}(h_z) =  -1.933 h_z^2  + 4.152 h_z^4 + \dots  \, .
\label{eq:EvacPert}
\end{equation}
One can also use the same $\sigma$ four-point function to compute the $O(h_z^2)$ correction to the state that corresponds to the $\sigma$ operator in the CFT limit:
\begin{equation}
m_1 R = \Delta_\sigma + 2.620 h_z^2 + \dots\, .
\label{eq:E1pert}
\end{equation}

Another approach one might try is to compute the spectrum in the $\epsilon = 4-d$ expansion at next-to-leading order, or a large $N_s$ expansion in the $O(N_s)$ model.
For the $\epsilon$ expansion with $N_s=1$,
\begin{equation}
\begin{aligned}
V(\phi) &= \frac{\lambda}{4!} \phi^4 + \frac{h_z}{V_d} \frac{1}{\cal N^{1/2}} \phi,
\end{aligned}
\end{equation}
where ${\cal N}$ is chosen so that the two point function of $\frac{1}{\cal N^{1/2}} \phi$ in position space is $\sim x^{-2\Delta_\phi}$; at leading order, ${\cal N} = \frac{\Gamma \mleft(\frac{d}{2}-1\mright)}{4\pi ^{d/2}}$ is just the normalization from the tree-level propagator. 
The fixed point coupling at leading order is 
\begin{equation}
 \lambda_* =  \frac{16 \pi^2}{3}\epsilon.
\end{equation}

Choosing $\phi$ to minimize $V(\phi)$, one obtains the  vacuum energy density ${\cal E}_{\rm vac}=  V(\phi)$ and one-particle mass:
\begin{equation}
\begin{aligned}
\frac{{\cal E}_{\rm vac}}{h_z^{\frac{d}{d-\Delta_\phi}} } &= - \frac{3^{5/3}}{8\pi^2 \epsilon^{1/3}}  \stackrel{\epsilon=1}{\approx}  -0.079, \\
\frac{m_1}{h_z^{\frac{1}{d-\Delta_\phi}}} &= 2^{1/2} 3^{1/6} \epsilon^{1/6} \stackrel{\epsilon=1}{\approx} 1.70.
\label{eq:EpsExpansionLO}
\end{aligned}
\end{equation}

One can include higher order corrections in the $\epsilon$ expansion; at the next order, the potential should include the one-loop Coleman-Weinberg potential as well.  However, without Borel-resummation, including such higher order terms requires a certain amount of unjustified optimism.

The calculation in the limit of large-$N_s$ of the $O(N_s)$ model can be found, for instance, in~\cite{Brezin:1972se}.  Unfortunately, neither approximation sheds much light on the two-particle binding energy in Ising Field Theory.  In the $\epsilon$ expansion, for $d>3$ the interaction in the nonrelativistic limit is irrelevant and consequently there is a minimum size for the coupling below which no bound state forms.  By definition, the coupling in the $\epsilon$ expansion is perturbative and therefore always below any such critical value.  In the large-$N_s$ approximation, the $(N_s-1)$ degenerate `transverse' modes are lighter than the radial mode, so a weakly-bound state of radial modes would not be kinematically stable in any case.

\section{Fuzzy sphere Setup}
\label{sec:FSsetup}
Let us review the FS setup for realizing the Ising CFT, which will be our starting point for deformation.  The review will be brief, mostly to illustrate our conventions.  For a more complete discussion, see the original work~\cite{Zhu:2022gjc}, or for example, our previous paper~\cite{Fardelli:2026zas}.

The FS is a system of two-flavor interacting non-relativistic fermions $\psi_i$ on a sphere with radius $R$ and in the presence of a magnetic monopole with total flux $4\pi s$ ($2s \in \mathbb{N}$).  In this setting,  energy eigenstates get quantized into spherical Landau levels.  Restricting to the lowest Landau Level (LLL), which has $2s+1$ degeneracy,  the fermions can be expanded into a complete orthonormal basis of monopole harmonics $\Phi_m$:
\twoseqn{
\psi_i(\Omega)&=\frac{1}{R}\sum_{m=-s}^s \Phi_m(\Omega)\, c_{m,i}\, , \qquad \qquad (i=\uparrow,\,\downarrow)\, ,
}[]
{
\Phi_m(\Omega)&=\sqrt{\frac{(2s+1)!}{4\pi (s+m)!(s-m)!}}e^{i m\phi} \cos^{s+m}\left(\frac{\theta}{2}\right)\sin^{s-m}\left(\frac{\theta}{2}\right)\,.
}[][]

We restrict to the set of states at half-filling,  i.e.~states where the number of fermions is $N=2s+1$. The Hamiltonian reads
\eqna{
H&=R^2\int d^2\Omega\,  \CH\, , \\
\CH&=\sum_{n=0,1}\lambda_n \left( n_0 \frac{{\nabla}^{2n}}{R^{2n}} n_0- n_z \frac{{\nabla}^{2n}}{R^{2n}} n_z \right)-h_x n_x\, ,
}[FSH]
where 
we have defined
\eqna{
n_i=\psi^\dagger \sigma_i \psi\, ,
}[niDef]
with  $\sigma_0=\mathds{1}$ and $\sigma_{x, y,z}$ the usual Pauli matrices.  There is a $\mathbb{Z}_2$ symmetry that flips the internal spin $\sigma^z \rightarrow - \sigma^z$ and a spacetime parity symmetry $\CP$, which acts within the space of $\ell_z=0$ states at half-filling by swapping filled and unfilled modes.

The parameters in the Hamiltonian, $\lambda_n$ are dimensionful and therefore their proper choice at criticality depends on introducing a regulator, which in our case is the truncation to the LLL.  Thus, in practice, we do not work with $\lambda_n$, but rather the dimensionless, so-called, Haldane potential parameters, $V_n$, which directly control terms in the Hamiltonian in terms of the LLL orbital creation/annihilation operators (i.e. the set $c_{m,i}, c_{m,i}^\dagger$).  These parameters are related to the $\lambda_n$'s as
\eqna{
V_0 &\equiv \frac{(2s+1)^2}{2\pi(4s+1)} \left( \frac{\lambda_0}{R^2}-s\frac{\lambda_1}{R^4}\right),  \\
V_1&\equiv  \frac{s(2s+1)^2}{2\pi(4s-1)}\frac{\lambda_1}{R^4}. \\ 
}[FPparameters]
In our previous work~\cite{Fardelli:2024qla},  we showed that, at $N=12$, the choice $V_0=4.825, V_1=1, h_x^c=3.158$ approximately corresponds to setting to zero the Wilson coefficients $g_{\CO}$s for $\CO=\epsilon$ and $\epsilon^\prime$ in the expansion near criticality of (\ref{eq:HFS-EFT}).  In this paper, we will continue to use this choice as a definition of the critical FS model.\footnote{Note, however, that these parameter values actually drift slightly with 
$N$, as shown in~\cite{Lauchli:2025fii}.}

Once one has tuned the microscopic parameters of the Fuzzy Sphere model to its critical point, Ising Field Theory is obtained by adding operators to the FS Hamiltonian that flow in the IR to one or both of the two relevant deformations, $\epsilon$ and $\sigma$, of the 3d Ising model.  The numeric analysis will generally be more efficient if the microscopic operators are as closely aligned as possible with the CFT operators one wants to study.   For the case of $\sigma$ that we are interested in,   we add to the Hamiltonian density in~\eqref{FSH} a term that explicitly breaks the $\mathbb{Z}_2$ symmetry
\eqna{
\CH\to \CH+ h_{z, \text{FS}} \lsp  n_z(\Omega)\, .
}[]
The microscopic operator $n_z$ admits an expansion in CFT operators of the form 
\eqna{
n_z(\Omega)=a_{\sigma} \sigma+a_{\partial^2_0  \sigma}\partial^2_0  \sigma+a_{\nabla^2  \sigma}\nabla^2  \sigma+a_{\CR \sigma}\CR  \sigma+\cdots\, , 
}[sigmaExp]
where the coefficients scale with the UV cutoff as
\eqna{
a_{\CO}\sim \left( \frac{1}{\Lambda_{\rm UV} }\right)^{\Delta_{\CO}}\, .
}[aSigma]
In~\cite{Hu:2023xak,Fardelli:2026zas},  it was shown explicitly that $n_z$ has indeed its largest overlap with $\sigma$ rather than with any of the other operators,  the latter being suppressed by the appropriate powers of the cutoff.\\
In section~\ref{sec:FSresults}, we will further reabsorb the dependence on the normalization $a_\sigma$  into a rescaling of $h_{z,\text{FS}}$, allowing us to directly compare the resulting coupling to the conventional field theory one in~\eqref{eq:IFT-Ham}.

\section{Hierarchy of scales and EFT Corrections} 
\label{sec:Hierarchy}
\begin{figure} \centering
\begin{tikzpicture}
\useasboundingbox (-6,-3) rectangle (6,3);
 \draw [line width=1.5pt,
             arrows = {-Stealth[length=4mm]}] (-6,-3) -- (-6,3);
\node at (4,0) { \includegraphics[width=0.4\textwidth]{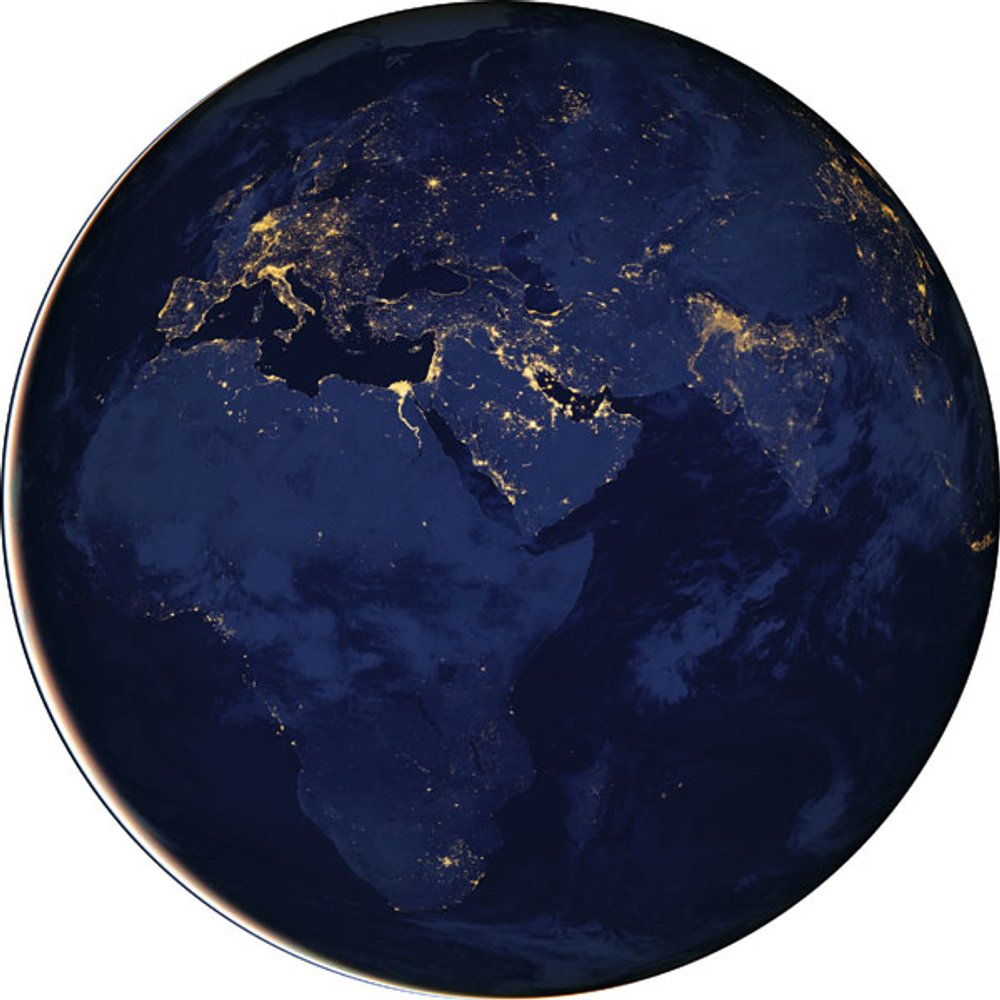}};
\node[left=0.2cm] at (-6,2.8) {$E$};
\draw[thick] (-6.25,2.4)--(-5.75,2.4);
\draw[thick] (-6.25,-2.8)--(-5.75,-2.8);
\draw[thick] (-6.25,0.5)--(-5.75,0.5);

\foreach \w/\op in {6/0.15, 4.5/0.25, 3/0.4, 1.5/1}{
  \draw[line width=\w pt, line cap=round, opacity=\op, orange]
        (-6,0.7) -- (-6,2.2);
}

\foreach \w/\op in {6/0.15, 4.5/0.25, 3/0.4, 1.5/1}{
  \draw[line width=\w pt, line cap=round, opacity=\op, greenl!80!black]
        (-6,0.3) -- (-6,-2.6);
}
\node[right=0cm] at (-5.75, 2.4) {$\Lambda_{\rm UV}\sim \frac{\sqrt{N}}{R}$};
\node[right=0cm] at (-5.75, 0.5) {$m_{\rm gap}\sim \frac{h_z^{\frac{1}{3-\Delta_{\sigma}}}}{R}$};
\node[right=0.2cm, orange] at (-5.75, 1.45) {3d Ising CFT};
\node[right=0.2cm, greenl] at (-5.75, -1.15) {massive gapped};
\node[right=0cm] at (-5.75, -2.8) {$\frac{1}{R}$};
\draw[thick, white!20!blul,-Stealth] (4.2,3.5)--(7.5,3.5);
\draw[thick, white!20!blul] (0.5,3.5)--(3.8,3.5);
\node[white!20!blul] at (4.0,3.5) {\small ${R}$};
\draw[thick, white!20!greenl] (3.6,0.5)--(4.9,0.5);
\draw[thick, white!20!greenl] (4.9,0.45)--(4.9,0.55);
\draw[thick, white!20!greenl] (3.6,0.45)--(3.6,0.55);
\node[white!20!greenl, below=0cm] at (4.25, 0.5) {\small $\xi\sim \frac{1}{m_{\rm gap}}$};
\draw[thick, white!20!orange] (2.55,1.3)--(2.8,1.3);
\draw[thick, white!20!orange] (2.55,1.27)--(2.55,1.33);
\draw[thick, white!20!orange] (2.8,1.27)--(2.8,1.33);
\node[white!20!orange, above=0cm] at (2.675,1.3) {\footnotesize $\ell_{\rm UV}\sim \frac{R}{\sqrt{N}}$};
\end{tikzpicture}\caption{Schematic view of hierarchy of scales involved.}
\label{fig:Hierarchy}
\end{figure}
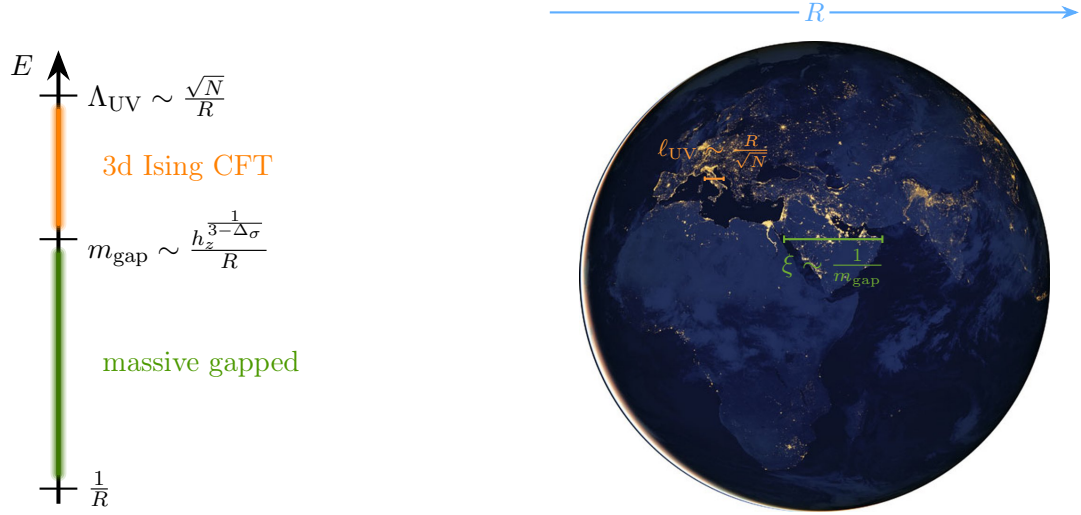

Our main theory of interest -- Ising Field Theory in flat space with a magnetic deformation -- has only a single scale, set by the coefficient $h_z$ of the deformation.  By contrast, our Fuzzy Sphere setup has additional scales, which modify the theory in both the UV and in the IR, and the main challenge of obtaining precise results is that we have to extract the behavior of the theory over an intermediate range of scales.   In this section, we will discuss the physics of these additional scales and how to parameterize their effects so that they can be removed efficiently.  

The basic situation is sketched in Fig.~\ref{fig:Hierarchy}.  The UV scale, $\Lambda_{\rm UV} \sim \frac{\sqrt{N}}{R}$, is the scale associated with the Fuzzy Sphere regulator itself.  This scale is conceptually similar to the inverse lattice spacing in a lattice microscopic model, where it is clear that the continuum QFT description completely breaks down.  The Fuzzy Sphere regulator does not break translations on the sphere, but by keeping on the Lowest Landau Levels (LLL) of the non-relativistic fermion, it smears out the canonical (anti-)commutation relations of the fermion over a length of parametric size $\Lambda_{\rm UV}^{-1}$.  Moreover, even when tuned to the critical point, the emergent IR conformal symmetry completely breaks down at scales above $\Lambda_{\rm UV}$ \cite{Fardelli:2024qla}.  

In the language of RG trajectories, the Fuzzy Sphere model only has a chance to start to approach the 3d Ising CFT below the scale $\Lambda_{\rm UV}$.  Even below the scale $\Lambda_{\rm UV}$, the theory can never perfectly reach the 3d Ising CFT, because of irrelevant deformations coming from the UV that are suppressed by powers of the cutoff $\Lambda_{\rm UV}$.  At best, if one studies quantities far below the cutoff, the theory can become very well-approximated by the 3d Ising CFT.  

On the other hand, the fact that we take space to be a sphere means that there is a maximum length scale available, namely the size of the sphere.  Consequently, it is not possible to study the theory at energies below $1/R$, which in turn means that irrelevant operators coming from the microscopic description are suppressed at most by powers of $\Lambda_{\rm UV} R \sim \sqrt{N}$.  The most important of these come from the lowest-dimension operators in the Hamiltonian that have not been explicitly tuned away.  In most of the fuzzy sphere literature, one chooses microscopic parameters that tune the coefficients of $\epsilon$ and $\epsilon'$ to be very small \cite{Fardelli:2024qla}, and then the leading irrelevant operator is the spin-4 operator $C_{\mu_1\cdots \mu_4}$ with dimension $\Delta_C \approx 5.02$.  By power-counting, these corrections are approximately linear in $1/N$:
\begin{equation}
\cal H \supset \frac{1}{\Lambda_{\rm UV}^{\Delta_C-3}} \int d^2 x \ C^{0000}, \qquad  \frac{1}{\Lambda_{\rm UV}^{\Delta_C-3}} \propto \frac{1}{N^{1.01}}.
\label{eq:HamIrrel}
\end{equation}
These are sometimes called `finite volume corrections', but in EFT terms it is more helpful  to think of  them as UV corrections.  The reason this distinction is worth emphasizing is that once we start to consider Ising Field Theory, we will have yet another scale, which is the scale $\mu \equiv h_z^{\frac{1}{3-\Delta_\sigma}}$ set by the relevant deformation.  Once this additional scale is present, it is possible to have corrections from irrelevant operators like (\ref{eq:HamIrrel}) that do not vanish in the infinite volume limit, but instead are merely suppressed by powers of $\mu/\Lambda_{\rm UV} \sim \mu R/\sqrt{N}$.  The main challenge of studying Ising Field Theory on the Fuzzy Sphere is that we want to take $N$ large and $\mu R$ large so that $\mu$ is small compared to the UV scale $\Lambda_{\rm UV}$ but large compared to the IR scale $1/R$:
\begin{equation}\label{eq:mulimits}
\frac{\mu}{\Lambda_{\rm UV}} \sim \frac{\mu R}{\sqrt{N}} \ll 1 \qquad  \textrm{ and } \qquad \mu R  \gg 1,
\end{equation}
which is manifestly only possible at very large $N$.   The right image in Fig.~\ref{fig:Hierarchy} shows somewhat more poetically how  the scale $\mu$, which sets the correlation length $\xi \sim \mu^{-1}$, is squeezed between the IR and the UV.  

There are additional UV effects that we have not included because they are subleading in $1/N$.  One such effect is due to the fact that even though the coefficient of $\epsilon$ is subtracted off by tuning when $h_z=0$, once $h_z$ is turned on the fuzzy sphere regulator induces additional contributions to $\epsilon$ from the $\sigma \times \sigma$ OPE.  These can be understood in conformal perturbation theory with a UV cutoff, and roughly can be thought of as a loop effect.  By simple power counting, at $O(h_z^2)$ the induced coefficient of $\epsilon$ is parametrically $O(h_z^2/\Lambda_{\rm UV}^{(2(3-\Delta_\sigma)-(3-\Delta_\epsilon))}) \sim O(h_z^2/\Lambda_{\rm UV}^{3.4})$.

In practice, we will take $N$ as large as possible and use extrapolations to effectively make it much larger.  At each value of the coupling $h_z$, we will extrapolate as a function of $N$ to get an approximate $N=\infty$ result.  However, this extrapolation only works for couplings such that $\mu R/\sqrt{N}$ is small, so that the corrections from (\ref{eq:HamIrrel}) are under control.  That in turn limits how large we can make $\mu R$.  Consequently, even with $\Lambda_{\rm UV}$ taken to infinity, we still  have to consider the finite volume energies from the introduction (\ref{eq:FiniteVolumeDimAnalysis}):
\begin{equation}
E_{\rm vac} =  \mu^3 R^2 f_0(\mu  R), \qquad m_i  = \mu f_i(\mu R) \,.
\end{equation}
It is again useful to consider the EFT description of the corrections to the large $\mu$ limit.  The effective Hamiltonian for the vacuum energy is a cosmological constant plus higher-order curvature terms (powers of the Ricci scalar $\CR$):
\begin{equation}
H = \int d^2 \Omega \left( {\cal E}_{\rm vac}^{(R=\infty)}  +  \mu^3 \sum_{n=1}^\infty c_n \left( \frac{\CR}{\mu^2} \right)^n + \dots \right),
\end{equation}
where the ``UV scale'' suppressing the curvature terms is $\mu$ since this is the scale of the gap, below which there are no more degrees of freedom.  In our fits, we will include only the flat-space vacuum energy density and a leading correction. In units where the radius $R=1$,
\begin{equation}
\frac{E_{\rm vac}}{\mu^3} \approx A_0 + \frac{\#}{\mu^2}.\,. 
\end{equation}
For the gap, we can use the following effective Lagrangian for the one-particle state:
\begin{equation}
\CL = \frac{1}{2} \left( \dot{\phi}^2 - (\nabla \phi)^2 - m^2_{(R=\infty)} \phi^2 - \sum_{m=1}^\infty \sum_{n=0}^\infty c_{m,n} \left( \frac{\CR}{\mu^2}\right)^m  \phi \left( \frac{\nabla^2}{\mu^2} \right)^n \phi \right). 
\end{equation}
Note that the sum over powers of $\CR$ begins at $m=1$, because the $m=0$ term has no dependence on the size of the sphere and therefore is fixed to be the Lagrangian of the flat space limit, which by assumption is a Lorentz invariant CFT deformed by a Lorentz-invariance preserving relevant deformation.  Lorentz invariance of the flat space limit fixes the relative coefficient of the $\dot{\phi}^2$ term and the $(\nabla \phi)^2$ term.  The $n=0$ terms are all completely equivalent to a volume-dependent mass.  As with the vacuum energy, we will fit the mass gap to a flat-space mass term plus a leading correction:\footnote{Such corrections were previously considered and tested in \cite{Hogervorst:2014rta}.}
\begin{equation}
\frac{m_i }{\mu} \approx A_i + \frac{\#}{\mu^2}.
\end{equation}

We will also use this effective description when we look at the spinning sector energies.  The lowest energy state at each nonzero spin $\ell$ is just a single particle state with spin.  Keeping only the first few terms in the effective description above, the dispersion relation at large $\mu R$ should be
\begin{equation}\label{eq:dispersion}
E_i(\ell) \approx E_{\rm vac} + \sqrt{m_i^2 +  \frac{b}{R^2}  \ell(\ell+1) + \frac{c }{R^2} \ell^2 (\ell+1)^2 +\dots}.
\end{equation}
The parameters $E_{\rm vac}, m_i, b, c, $ etc. all depend on $\mu R$.  Lorentz invariance in the flat space limit, where momentum $p \sim \ell/R$ is fixed as $R\rightarrow \infty$, implies that a large $\mu R$, $b(\mu R) \stackrel{\mu R \rightarrow \infty}{\rightarrow} 1$. 
Moreover flat-space Lorentz invariance forbids a $p^4/\mu^2$  term, therefore the leading quartic correction must carry a factor of $\CR/\mu^2$,  such that $c(\mu R) \stackrel{\mu R \gg 1}{\sim} (\mu R)^{-4}$, which is very small at large $\mu R$.  
Our Fuzzy Sphere data is limited to a small number of different values of the spin $\ell$, $0 \le \ell \le 4$ (errors increase rapidly with $\ell$, because of the enhancement of higher-derivative terms $\nabla^2 \sim - \ell(\ell+1)/R^2$), and including too many fit parameters rapidly leads to overfitting.  We have found that fitting $E_{\rm vac}, m_1, b$ and $c$ leads to results for $E_{\rm vac}$ and $m_1$ that do not match well with the direct extraction of $E_{\rm vac}$ and $m_1$ from the first and second lightest energies of the $\ell=0$ sector. However, if we fix $b=1$, which we expect to be a good approximation at large $\mu R$ by the argument above, then we obtain much better agreement, as we will show in later sections.

 \section{Fuzzy sphere Results} \label{sec:FSresults}
 \begin{figure}[t!]\centering
 \includegraphics[width=0.6\textwidth]{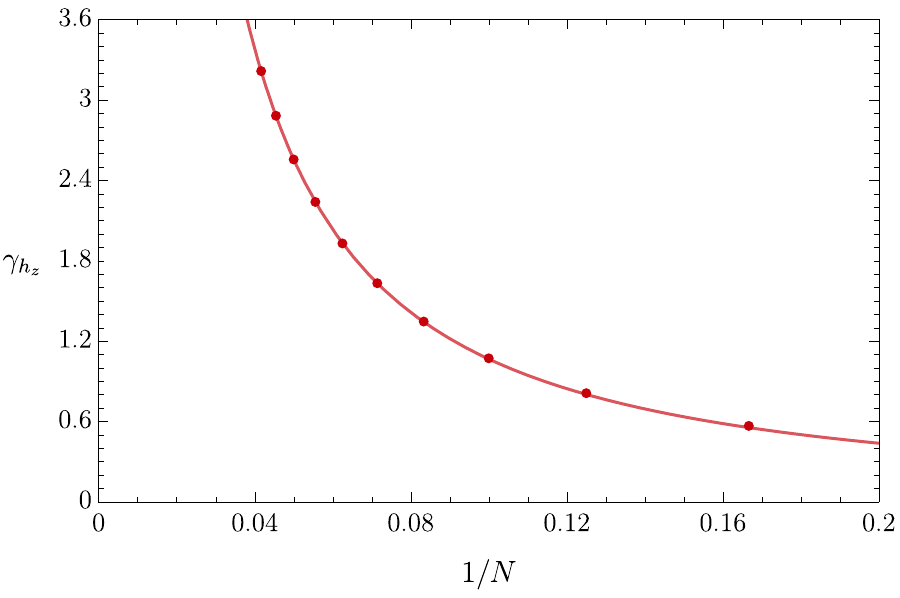}\caption{ $\gamma_{h_z}$ rescaling factor converting the fuzzy sphere coupling to the conventional field theory definition.  The solid line corresponds to the fit $N^{\frac{3-\Delta_{\sigma}}{2}}\left(0.06	-\frac{0.02}{N}\right)$.}
 \label{fig:CouplingRescalingFactor}
 \end{figure}
\begin{figure}\centering
 \includegraphics[width=0.6\textwidth]{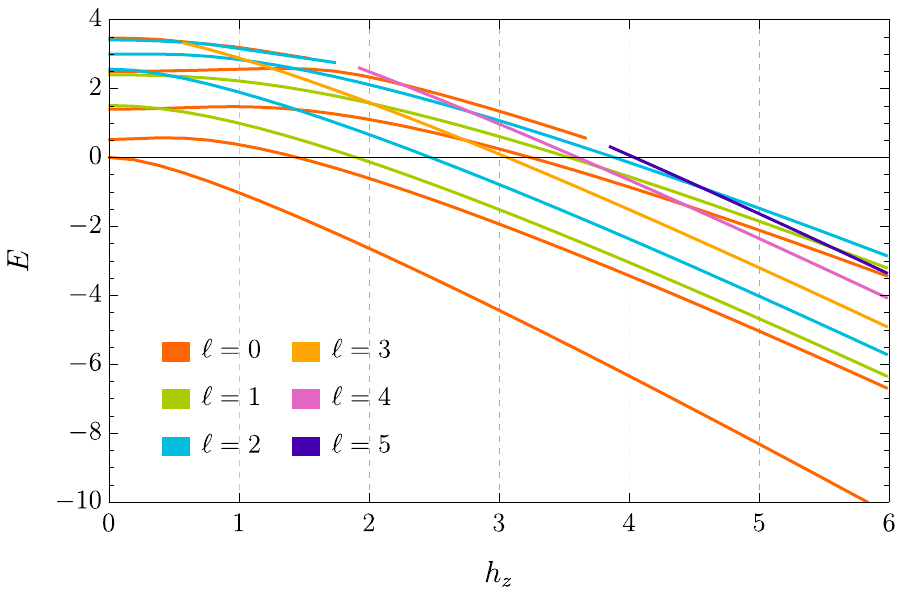}\caption{Spectrum obtained from ED at $N=16$ keeping the 10 lowest eigenvalues.  The spectrum has been normalized by requiring $E_{\rm vac}(h_z=0)=0$ and $E_1(h_z=0)=\Delta_\sigma$.  $h_{z,\text{FS}}$ has been rescaled to match $h_z$ according to~\eqref{eq:hzTCSAtoFS}.}
 \label{fig:RawSpectrum}
 \end{figure}
In this section, we will present our results for the spectrum of Ising Field Theory from the Fuzzy Sphere regulator.  When determining the spectrum, we will always first subtract the energy of the CFT vacuum and rescale the overall size of the Hamiltonian so that the energy spectrum when the deformation is turned off ($h_z =0$) matches the CFT spectrum of scaling dimensions.\footnote{In practice, we rescale the Hamiltonian so the lowest-energy $\mathbb{Z}_2$-odd state matches the dimension $\Delta_\sigma$. }  We also rescale the coupling $h_{z,\rm FS}$ in the microscopic Hamiltonian in order to match the conventional Field Theory normalization for the coefficient $h_z$ of the primary operator $\sigma$ in the IR Hamiltonian.  The simplest way to match this normalization is to compare the vacuum energy at small $h_{z, \text{FS}}$ and rescale it to match the conformal perturbation theory prediction (\ref{eq:CPTQuadVac}).  The rescaling factor,
\begin{equation}\label{eq:hzTCSAtoFS}
h_z = \gamma_{h_z} h_{z, \rm FS},
\end{equation}
is shown in Fig.~\ref{fig:CouplingRescalingFactor} for several values of $N$.  The dependence on $N$ agrees with the expectation
\eqna{
\gamma_{h_z}\sim N^{\frac{3-\Delta_\sigma}{2}}\left( 1+O\left( \frac{1}{N}\right)\right)\, , 
}[]
where the $\Delta_{\sigma}$-dependent part follows from the operator expansion in~\eqref{sigmaExp} together with~\eqref{aSigma}, and the remaining dependence comes from the rescaling of $H$. The $1/N$-corrections come from subleading terms in~\eqref{sigmaExp} and  the precise relation between  $R$ and $N$.

In the rest of the paper, all spectra up to $N=16$ are obtained by Exact Diagonalization (ED), while those at larger $N$ using Density Matrix Renormalization Group (DMRG), as implemented in the Julia package FuzziED~\cite{Zhou:2025liv}.
As an illustration,  Fig.~\ref{fig:RawSpectrum} shows the raw spectrum for the lowest 10 eigenvalues at $N=16$. Because angular momentum is exactly preserved by the FS, eigenvalues for states with different angular momenta can cross each other, so that the spin of the tenth-lowest-eigenvalue can change when two of these lines cross.  An important consequence of this fact is that there is an interesting, if unfortunate, obstruction to obtaining several low-energy {\it scalar} eigenvalues. The obstacle is that to obtain, say,  the first $n$ lowest scalar eigenvalues generally requires obtaining more -- sometimes many more -- than $n$ total (scalar $+$ spinning) eigenvalues.  At large coupling (equivalently, large volume), the obstruction is a dynamical one, because of the emergence of Lorentz invariance in the large volume. Concretely, approximate Lorentz invariance implies that many spinning states will essentially just be boosted version of the lightest single particle state, and therefore remain relatively light at large coupling.  We will study this phenomenon, as well as other relations between the spectra in different spin sectors, in detail in section~\ref{sec:spinningSectors}, but we will first focus on the spin-0 subsector.
\begin{figure}\centering
 \includegraphics[width=0.49\textwidth]{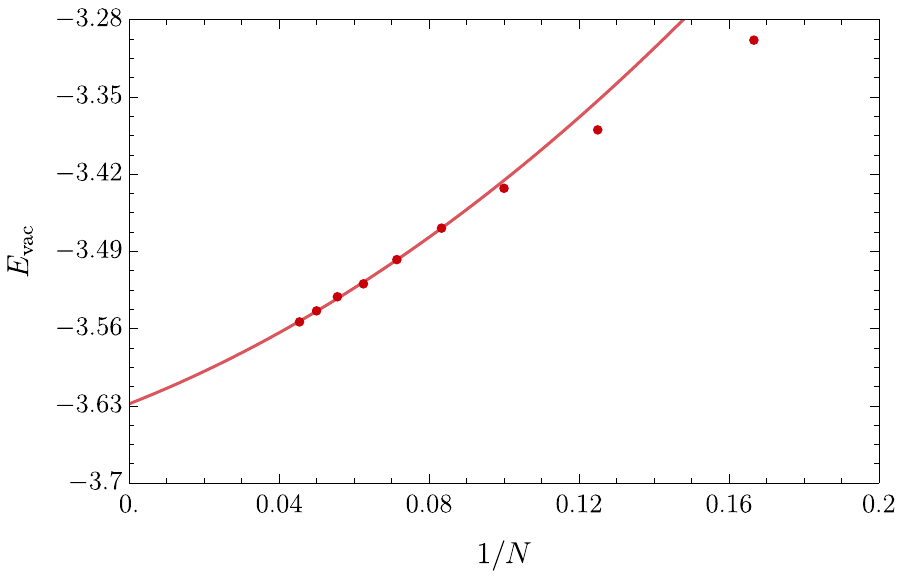}
  \includegraphics[width=0.49\textwidth]{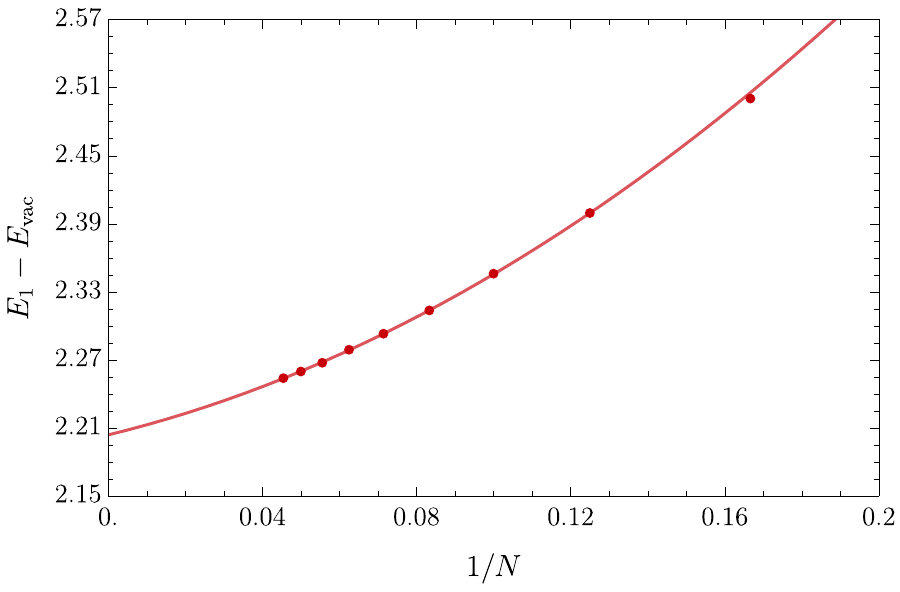}\caption{$N$-dependence of $E_{\rm vac}$  (\textit{left}) and $E_1-E_{\rm vac}$ (\textit{right}) at $h_z=2.5$.  Solid lines are fits made using points 12 to 22: for $E_{\rm vac}$ the fit corresponds to $-3.6+\frac{1.3}{N}+\frac{6.9}{N^2}$, while for $E_1-E_{\rm vac}$ it is $2.2+\frac{0.83}{N}+\frac{5.9}{N^2}$.}
 \label{fig:EvacOfNhz2}
 \end{figure}
 
Before doing so, however, it is useful to examine the dependence of the lowest energy levels on $N$ and to connect it to the discussion in Section~\ref{sec:Hierarchy}. In particular, at sufficiently large $N$, we expect the leading large-$N$ corrections to be controlled by irrelevant operators whose Wilson  coefficients have not been tuned  to zero, even when $h_z=0$ and whose contribution follows from~\eqref{eq:HamIrrel}.
In Fig.~\ref{fig:EvacOfNhz2}, we show the dependence of the vacuum energy $E_{\rm vac}$ and the energy gap $E_1-E_{\rm vac}$ on $N$, at fixed $h_z=2.5$, in the region where we expect definite scaling with $N$ to hold.  The data are well described by the fit
\eqna{
a+\frac{b}{N}+\frac{c}{N^2}
}[]
over the range of $N$ shown. More precisely, the leading correction predicted by~\eqref{eq:HamIrrel} would scale as $N^{-1.01}$. However, since several irrelevant operators with similar scaling dimensions can contribute, we find that fitting to integer powers of $1/N$ provides a more stable and convenient parametrization of the data, while capturing the behavior relevant for our purposes.

\subsection{Results at spin 0}
 \begin{figure}\centering
 \includegraphics[width=0.49\textwidth]{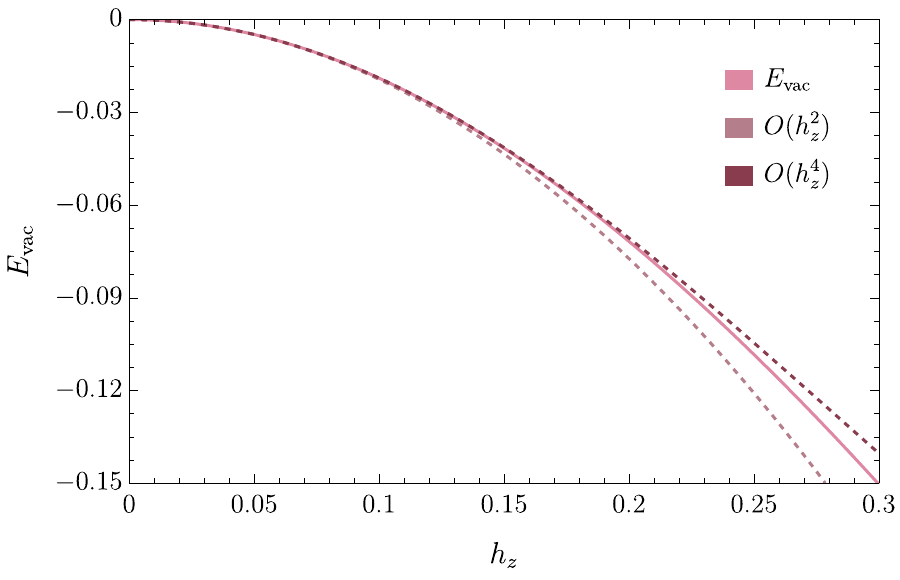}
  \includegraphics[width=0.49\textwidth]{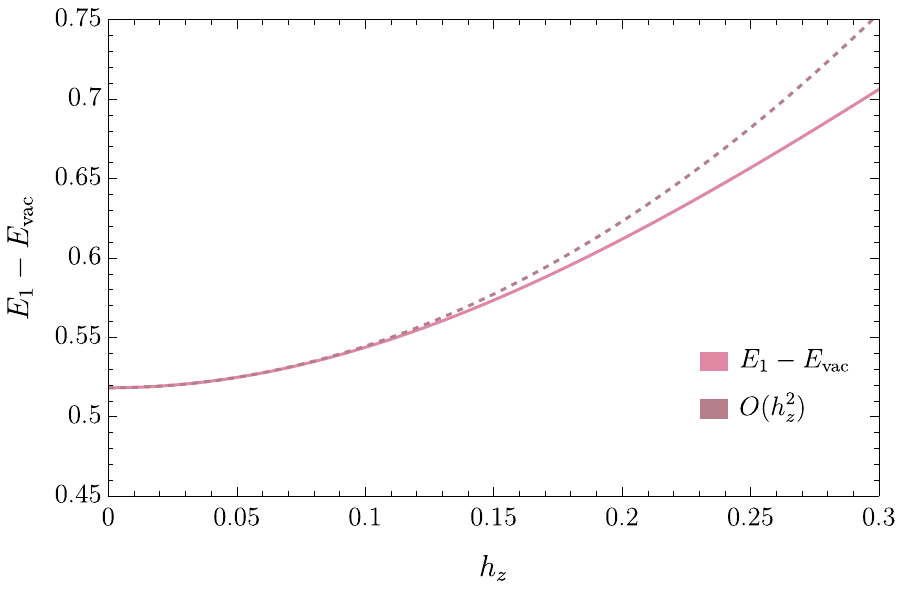}\caption{{\it Left:} Vacuum energy $E_{\rm vac}$ at small coupling $h_z$, compared with the quadratic $O(h_z^2)$ and quartic $O(h_z^4)$ approximations, $E_{\rm vac}(h_z) = -1.933 h_z^2 + 4.152 h_z^4$, from (\ref{eq:EvacPert}). {\it Right:} Scalar sector mass gap $E_1 - E_{\rm vac}$, compared with the $O(h_z^2)$ quadratic approximation  (\ref{eq:E1pert}).  }
 \label{fig:EvacPert}
 \end{figure}
\subsubsection*{Vacuum Energy}

\begin{figure}\centering
 \includegraphics[width=0.49\textwidth]{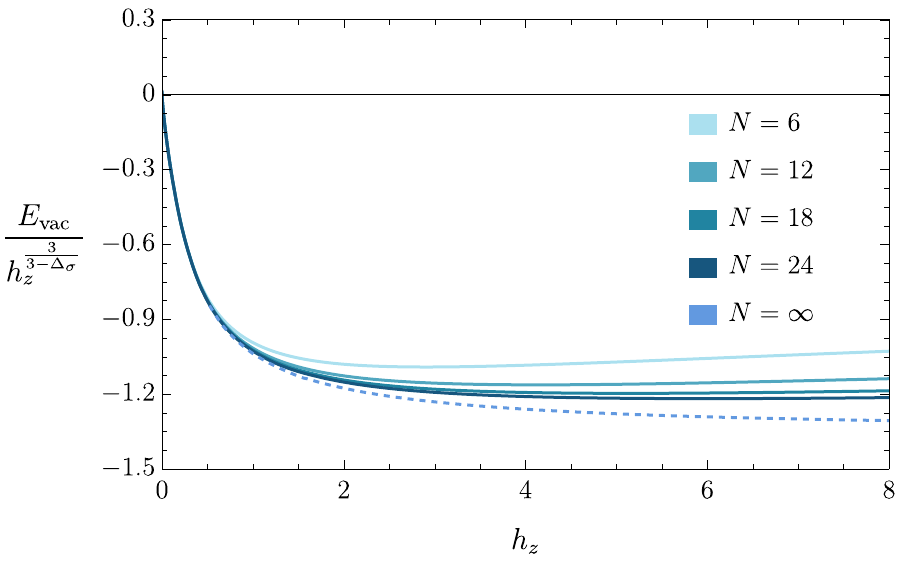}
  \includegraphics[width=0.49\textwidth]{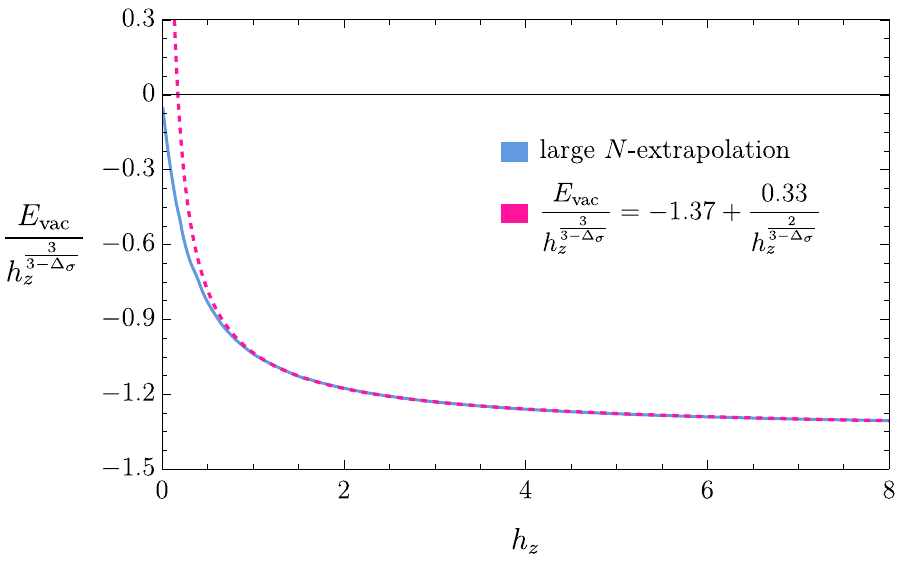}\caption{{\it Left:} Ground state energy in units of $\mu^3$ after rescaling of the spectrum and of $h_z$. $N\leq16$ is obtained by ED, DMRG otherwise. The $N\to\infty$ limit is done by fitting to $a+\frac{b}{N}+\frac{c}{N^2}$ for $N=14, 16, 18, \cdots, 24$. {\it Right:} Ground state energy in units of $\mu^3$ large $N$ extrapolation vs fit to curvature corrections.  At large $N$ the fit is $a+b/\mu^2$ and is done for $1<h_z<4$, with the result $\{a = -1.36972, b =0.336073\}$.  For comparison, the fit at $N=24$ for $1<h_z<3$ is $\{a=-1.3122, b =0.282027\}$. }
 \label{fig:EvacOfN}
 \end{figure}

First, consider the vacuum energy $E_{\rm vac}$.  At small $h_z$, we can compare it to the Conformal Perturbation Theory (CPT) calculation in (\ref{eq:EvacPert}). The quadratic $O(h_z^2)$ approximation is what we use to determine the relative normalization of the Fuzzy Sphere deforming operator and the Ising CFT $\sigma$ operator, and so fits by construction.

At larger values of the coupling, the vacuum energy should transition to the flat-space scaling $E_{\rm vac} \propto h_z^{\frac{3}{3-\Delta_\sigma}}$.  The raw results deviate from the flat-space scaling because of both UV effects and IR effects, coming from finite $N$ and finite $h_z$ respectively.  In Fig.~\ref{fig:EvacOfN}, we show the vacuum energy as a function of $h_z$ at various values of $N$.  At smaller values of $h_z$, the curves converge to their $N=\infty$ limit faster. 
To obtain a better estimate of the infinite $N$ result, at each value of $h_z$ we do a fit of the vacuum energy to the form $a+b/N + c/N^2$.  The $N=\infty$ extrapolated value is shown by the dashed curve in Fig.~\ref{fig:EvacOfN}.  

Even at $N=\infty$, there are finite volume corrections from the size of the sphere, so that the infinite volume scaling $E_{\rm vac} \sim h_z^{\frac{3}{3-\Delta_\sigma}}$ gets corrections at smaller $h_z$.  As discussed in section~\ref{sec:Hierarchy}, at large volume these corrections should be captured by local curvature terms, which produce contributions with inverse powers of $h^{\frac{2}{3-\Delta_\sigma}}$.  In Fig.~\ref{fig:EvacOfN}, we compare the infinite $N$ dependence with a fit to the large volume behavior including the leading curvature correction, given by
\begin{equation}
\frac{E_{\rm vac}}{\mu^3} = -1.37 + \frac{0.33}{\mu^2} , \qquad \mu \equiv h^{\frac{1}{3-\Delta_\sigma}}. 
\end{equation}
The infinite volume energy density is ${\cal E}_{\rm vac} \equiv \frac{E_{\rm vac}}{4\pi} \approx -0.109 \mu^3$, which can be compared with the leading order $\epsilon$ expansion prediction of (\ref{eq:EpsExpansionLO}). 
As a rough and  conservative estimate of the error in these numbers in the caption of Fig.~\ref{fig:EvacOfN}, we show the results for the fit to the leading curvature correction performed at $N=24$.
\subsubsection*{Mass Gap}
\begin{figure}\centering
 \includegraphics[width=0.49\textwidth]{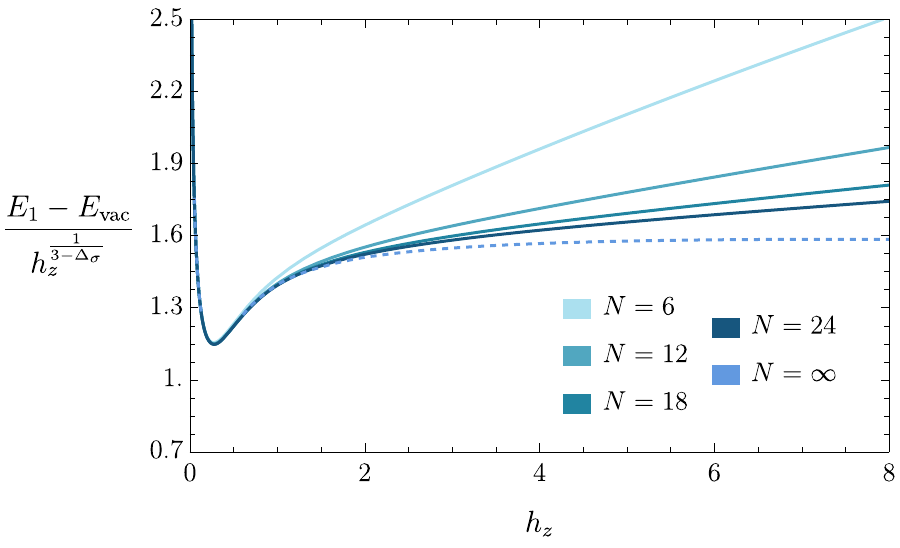}
  \includegraphics[width=0.49\textwidth]{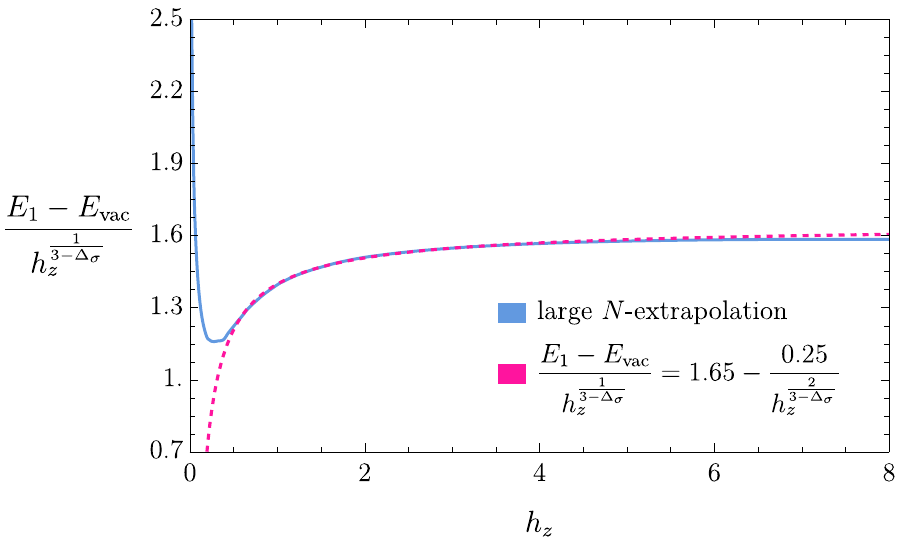} \caption{{\it Left:} Gap in units of $\mu$ after rescaling of the spectrum and of $h_z$. $N\leq16$ is obtained by ED, DMRG otherwise. The $N\to\infty$ limit is done by fitting to $a+\frac{b}{N}+\frac{c}{N^2}$ for $N=14, 16, 18, \cdots, 24$. {\it Right:} Gap in units of $\mu$ large $N$ extrapolation vs fit to curvature corrections.  At large $N$ the fit $a+b/\mu^2$ is done for $1<h_z<4$ and the result is $\{a = 1.65298, b =-0.254676\}$.  For comparison, the fit at $N=24$ for $1<h_z<3$ is $\{a=1.70914, b =-0.322984\}$. }\label{fig:E1inN}
 \end{figure}
Next, we consider the first excited state, which interpolates continuously from the $\sigma$ state in radial quantization at $h_z=0$ to the lightest one-particle state at large $h_z$.  Its energy at small coupling is shown in Fig.~\ref{fig:EvacPert} and compared to the $O(h_z^2)$ approximation (\ref{eq:E1pert}).  The $N=\infty$ value shown in Fig.~\ref{fig:E1inN} is again obtained by fitting the results for finite $N$ to a linear plus quadratic correction $a+b/N +c/N^2$.  Then, to obtain the flat space limit, we fit the $N=\infty$ curve as a function of $h_z$ to its large volume scale plus a curvature $\sim 1/\mu^2$ correction, with the result:
\begin{equation}
\frac{E_1 - E_{\rm vac}}{\mu} = 1.65 - \frac{0.25}{\mu^2},  \qquad \mu \equiv h^{\frac{1}{3-\Delta_\sigma}}. 
\end{equation}
For comparison, the leading order in the $\epsilon$ expansion (\ref{eq:EpsExpansionLO}) predicts that $(E_1 - E_{\rm vac})/\mu \approx 1.70$.   Also in this case,  an estimate of the error can be found in Fig.~\ref{fig:E1inN} from the result of the fit to  curvature correction  at $N=16$.

The most obvious qualitative feature of the gap as a function of $h_z$ is probably that it approaches its large volume limit from below.  One intuitive way to think about this correction is that the magnetic deformation is like a $\phi^3$ interaction (for instance, the two are connected by the $\epsilon$ expansion).  The contribution from a scalar loop, with a cubic coupling, on the sphere also generates a mass that is lowered by finite volume effects (see e.g. \cite{Fardelli:2026zas}).  
 \subsubsection*{Bound-state and first two-particle state}
 \begin{figure}\centering
 \includegraphics[width=0.49\textwidth]{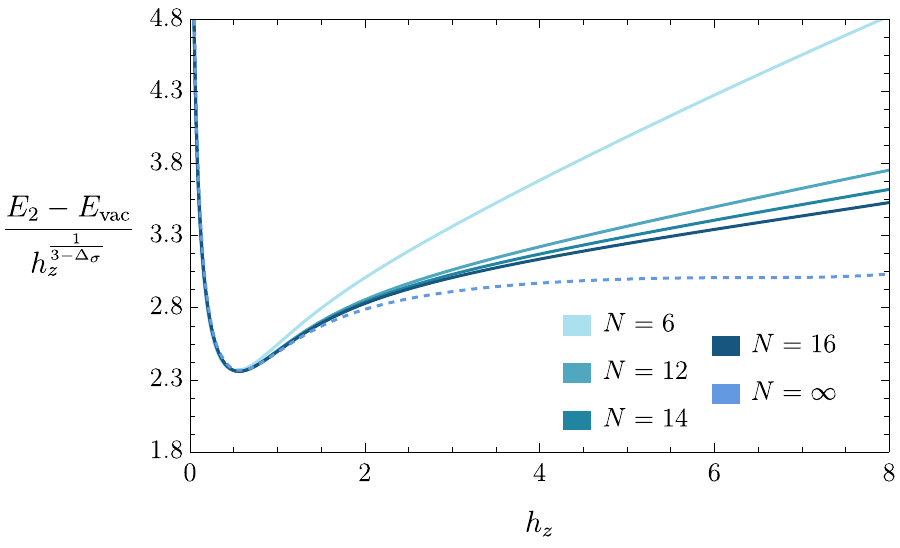}
  \includegraphics[width=0.49\textwidth]{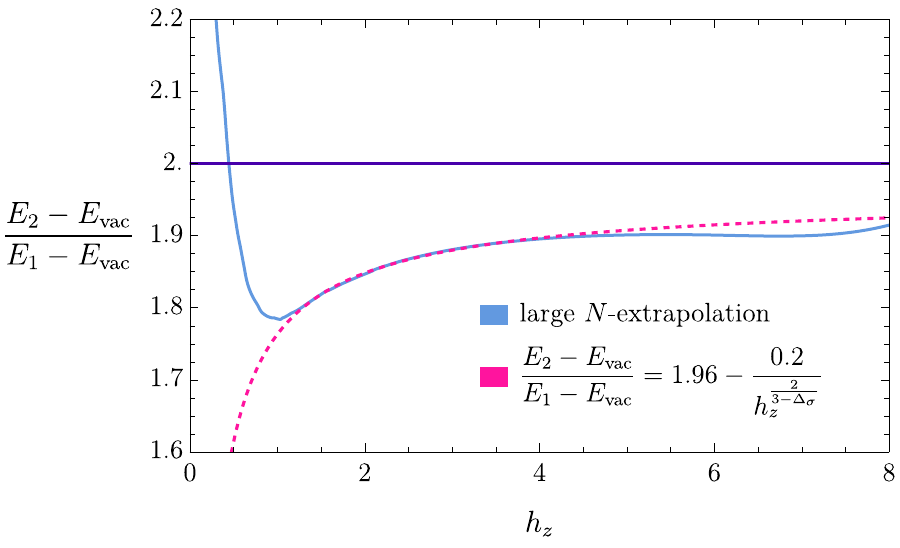}
  \caption{{\it Left:}  Bound state in units of $\mu$ after rescaling of the spectrum and of $h_z$.  Only ED for $6\leq N\leq16$. The $N\to\infty$ limit is done by fitting to $a+\frac{b}{N}+\frac{c}{N^2}$ for $N=10, 12,14,16$. {\it Right:} Bound state in units of $\mu$ large $N$ extrapolation vs fit to curvature corrections.  At large $N$ the fit is $a+b/\mu^2$ is done for $1<h_z<4$ and the result is $\{a = 1.96187, b = -0.198938\}$, for comparison, the fit at $N=16$ for $1<h_z<3$ is $\{a=1.93527, b = -0.161706\}$.}\label{fig:bound}
 \end{figure}

Finally, we consider the second and third excited states. At $h_z=0$,  these states can be identified, respectively, with the operator $\epsilon$ and the scalar descendant $\Box\sigma$. At large $h_z$, on the other hand, they are naturally interpreted as a stable bound state and a two-particle state. 
 In the left panel Fig.~\ref{fig:bound}, we show the energy difference $E_2-E_{\rm vac}$ in units of $\mu$ for different values of $N$. Unlike in the previous analysis, we only report ED results. Performing the corresponding analysis with DMRG is considerably more challenging. Although the state of interest is the second excited state within the scalar sector, as $h_z$ increases, spinning states can fall below it in the spectrum of the full Hamiltonian, as illustrated in Fig.~\ref{fig:RawSpectrum}. The scalar state is therefore pushed to higher energies in the full spectrum. Accessing such higher excited states is computationally expensive with DMRG, and the accuracy deteriorates as the number of targeted states increases. A useful improvement for future studies would therefore be to restrict the computation to the scalar sector directly in the fuzzy sphere setup, which would allow for a more systematic analysis of these higher scalar states.

In the right panel of Fig.~\ref{fig:bound}, we show the ratio of  $E_2-E_{\rm vac}$ to the mass gap,  extrapolated at large $N$. 
The ratio remains below the two-particle threshold, 
\eqna{
\frac{E_2-E_{\rm vac}}{E_1-E_{\rm vac}}<2\, ,
}[]
indicating that this state is kinematically stable against decay into two lightest particles and can therefore be interpreted as a bound state. To further isolate the flat-space limit, we proceed as before and we fit the large-$N$ extrapolated ratio as a function of $h_z$ to a constant plus the leading curvature correction,
\eqna{
\frac{E_2-E_{\rm vac}}{E_1-E_{\rm vac}}=1.96-\frac{0.2}{\mu^2}\,.
}[]
The extrapolated value is close to two,  indicating  a small binding energy.

The finite volume data (specifically, with $N$ extrapolated to $\infty$, but at finite volume in units of the IR scale $\mu$) are also consistent with the result of~\cite{Taylor:2026wan}.\footnote{In particular, our results for the spectrum at $N=12,14,16$ coincide with those in Fig. S6 (a) in~\cite{Taylor:2026wan}.} In particular at the value $h_z =1.625$ used in \cite{Taylor:2026wan},\footnote{This is the value of the coupling used to produce Table 1 in~\cite{Taylor:2026wan}. In their notation, $h_z =1.625$ is $\mu_{\rm there}=26$, since $h_z\approx  0.06 \mu_{\rm there}$; our $\mu \equiv h_z^{\frac{1}{3-\Delta_\sigma}}$ that we use in this paper is a different quantity from their $\mu_{\rm there}$.}  the mass ratio we obtain in our convention is
\eqna{
\frac{E_2-E_{\rm vac}}{E_1-E_{\rm vac}}(h_z=1.645)=1.827\,,
}[]
which should be compared  to $m_2=1.829(2)$ in~\cite{Taylor:2026wan}.

 \begin{figure}\centering
 \includegraphics[width=0.6\textwidth]{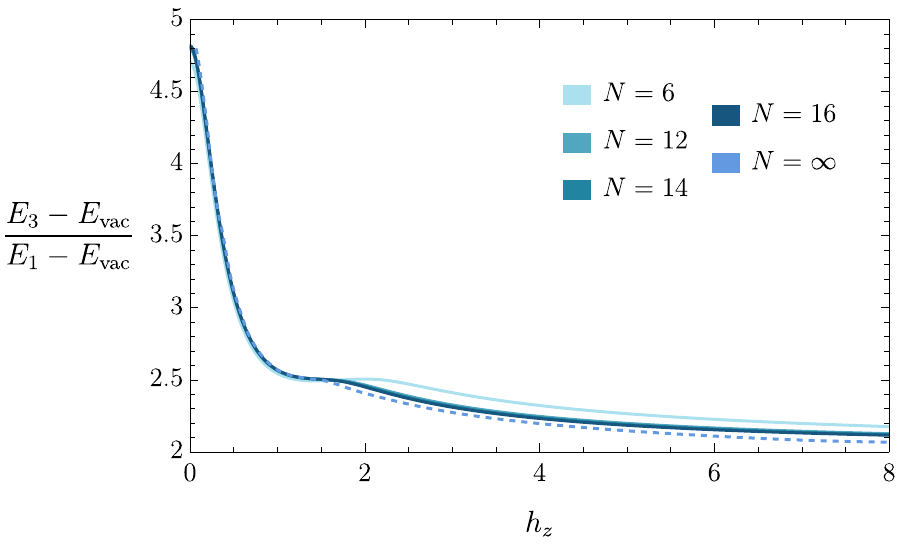}
 \caption{Energy of the third excited state in units of the mass gap for $6\leq N\leq 16$ using ED and $N\to \infty$ limit obtained by fitting to $a+\frac{b}{N}+\frac{c}{N^2}$ for $N=10,12,14,16$.  } \label{fig:twoparticle}
 \end{figure}
Finally Fig.~\ref{fig:twoparticle} shows the ratio of the third excited state to the energy gap.  Both the finite-$N$ results and the large-$N$ extrapolation lie above the two-particle threshold,  supporting its interpretation as a two-particle state.  Notice that  the ratio approaches 2 from above as $h_z$ is increased,  as expected for a scattering state of two particles carrying nonzero relative angular momentum.  The different behavior as a function of $h_z$ of this state, as compared to the bound state, is partly due to avoided level crossings between states in the scalar sectors.   This level crossing of the eigenstates makes a controlled fit to the curvature corrections unreliable, and we therefore do not attempt such a fit here.  Fig.~\ref{fig:twoparticle} is instead intended as a qualitative characterization of the fact that the scalar state of interest remains above the two-particle threshold.

 \subsection{Results from spinning sectors} \label{sec:spinningSectors}
   \begin{figure}\centering
 \includegraphics[width=0.6\textwidth]{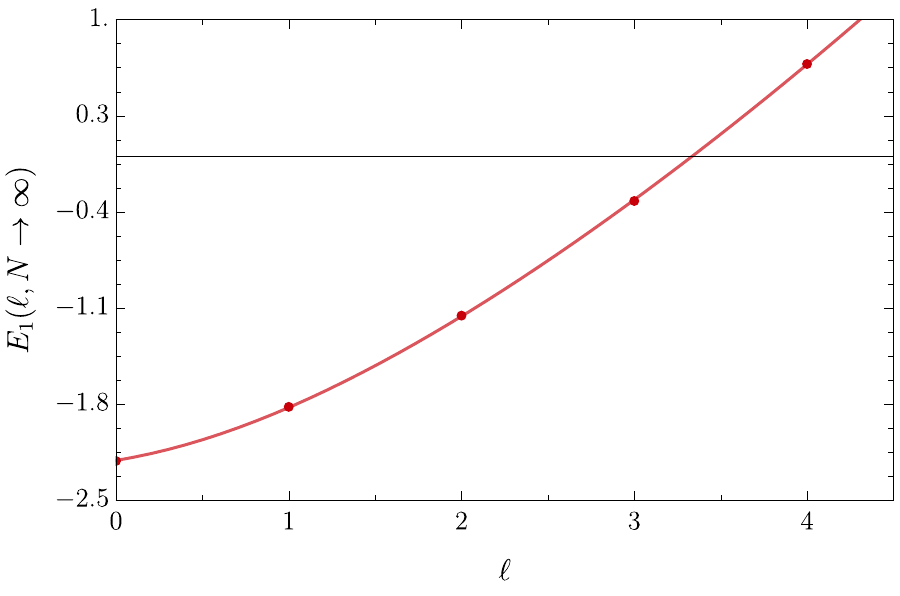}\caption{Large $N$ extrapolation of $E_1$ for different spin sectors $\ell$ at  $h_z=3$.  Solid line corresponds to the fit  $E_{1}(\ell, N\to \infty)=-{1.23}\, {h_z^{\frac{3}{3-\Delta_{\sigma}}}}+\sqrt{ (1.56\, {h_z^{\frac{1}{3-\Delta_{\sigma}}}})^2+\ell(\ell+1)+0.006\, \ell^2 (\ell+1)^2}$.
 }\label{fig:ellfits}
 \end{figure}
 
 In this section we show how the spinning  energy levels can be used to  extract information about the vacuum energy and 
 the mass gap.  Following the discussion in section~\ref{sec:Hierarchy}, 
we fit the first excited level at each angular momentum to
  \begin{align} \label{eq:ellfit}
 E_1(\ell)=E_{\rm vac}+\sqrt{m_{\rm gap}^2+\ell(\ell+1)+c\, \ell^2 (\ell+1)^2}\,, 
 \end{align}
 where we have set $R=1$ and fixed the coefficient of the $\ell(\ell+1)$ to one as discussed around~\eqref{eq:dispersion}.

To extract the spectrum, we select $J_z=\ell$ for $\ell=1,2,3,4$ on the fuzzy sphere, using  ED for $N\leq 16$ and DMRG up to  $N= 22$.  As in the scalar sector,  we rescale $h_{z, \text{FS}}$ to match the field-theory convention and normalize the spectrum at $h_z=0$.  As a first check that this setup is consistent,   Fig.~\ref{fig:ellfits} shows $E_1(\ell)$ at fixed $h_z=3$,  extrapolated to $N\to \infty$ for different values of $\ell$. The result is well described by the form in~\eqref{eq:ellfit}, with values of $E_{\rm vac}$ and $m_{\rm gap}$ consistent with those obtained independently from the scalar sector in the previous section.

\begin{figure}\centering
\begin{minipage}{0.48\textwidth}\centering
 \includegraphics[width=1\textwidth]{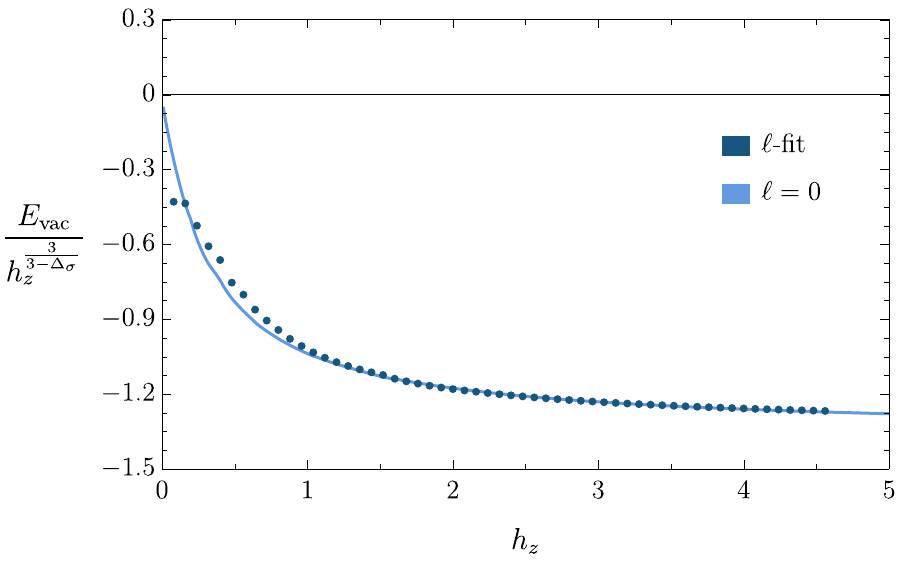}
 \end{minipage}
 \hfill
 \begin{minipage}{0.48\textwidth}\centering
 \includegraphics[width=1\textwidth]{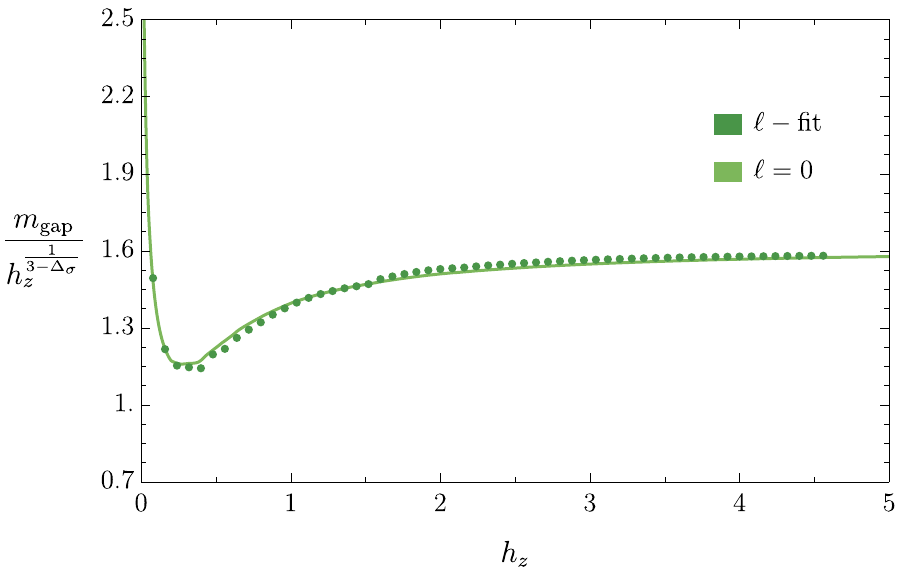}
 \end{minipage}
 \caption{Comparison of the vacuum energy and mass gap obtained from the scalar sector and from the spinning-sector analysis. For the latter, we first fit the spectrum as a function of $\ell$ as in~\eqref{eq:ellfit} for  each value of $N$ and then extrapolate to large $N$ using a quadratic ansatz in $1/N$ for $N=12,14,\cdots, 22$. }\label{fig:vacGapFromEll}
 \end{figure}

In Fig.~\ref{fig:vacGapFromEll},  we instead proceed in the opposite order: for each value of $N$ we first fit the spectrum as a function of $\ell$, including $\ell=0$, to obtain  $E_{\rm vac}^{\ell\text{-fit}}(N)$ and $m_{\rm  gap}^{\ell \text{-fit}}(N)$ and then we perform a quadratic fit  in $1/N$.\footnote{It is worth mentioning that the convergence with $N$ is, however,  faster in the scalar-sector analysis, since the higher-energy eigenvalues used in the spinning-sector fit are less accurately determined at finite $N$ than the lowest-lying states.} The resulting leading large-$N$ extrapolations  are compared with the corresponding quantities extracted directly from the scalar sector. We find very good agreement for both the vacuum energy and the mass gap.
Strictly speaking Lorentz invariance is expected to emerge only at large volume, so it is quite surprising that the agreement with the $\ell=0$ predictions extends also to small values of $h_z$.

All in all,  this agreement provides  a nontrivial consistency check. Although the $\ell=0$ energy $E_1(\ell=0)$ is included in the fit, the vacuum energy $E_{\rm vac}$ obtained from the scalar sector is never used as an input.  The spinning states therefore provide an independent determination of the vacuum energy and mass gap, and their agreement with the scalar-sector results confirms the consistency of the relation in~\eqref{eq:ellfit}.

 \subsection{Overlaps}
  \begin{figure}\centering
 \includegraphics[width=0.48\textwidth]{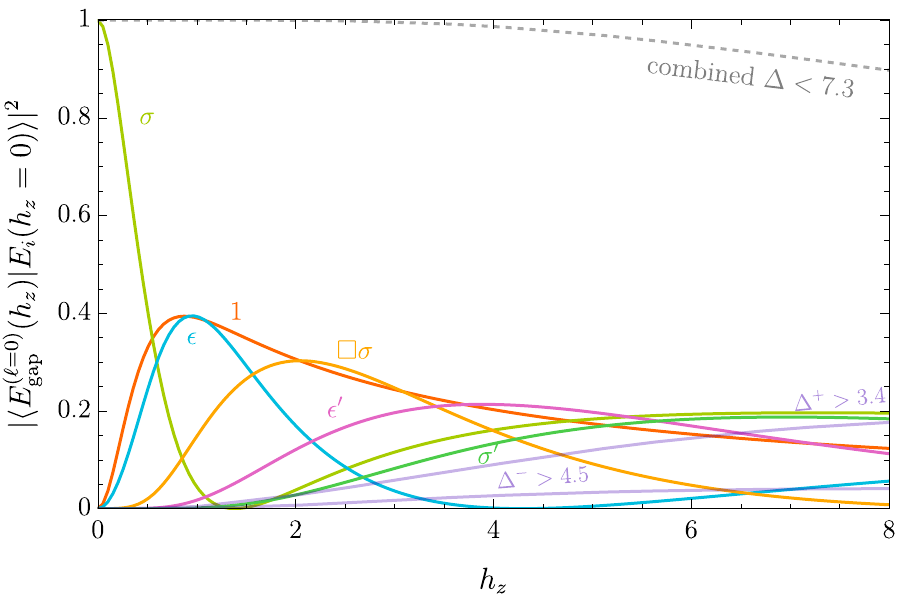}
 \caption{The CFT state content of the one-particle state for different values of $h_z$ in the scalar sector.  Purple lines indicate $\mathbb{Z}_2$-even (+) and $\mathbb{Z}_2$-odd (-) states which have not been clearly identified (see text).  The gray dashed line shows the total squared overlap captured by the CFT states included in the analysis, obtained by summing over the 14 scalar states among the 150 lowest-energy states retained at $h_z=0$. }
 \label{fig:overlapspin0} 
 \end{figure}
 
\begin{figure}
    \centering
    \begin{subfigure}[t]{0.48\textwidth}
        \centering
        \includegraphics[ width=1\textwidth]{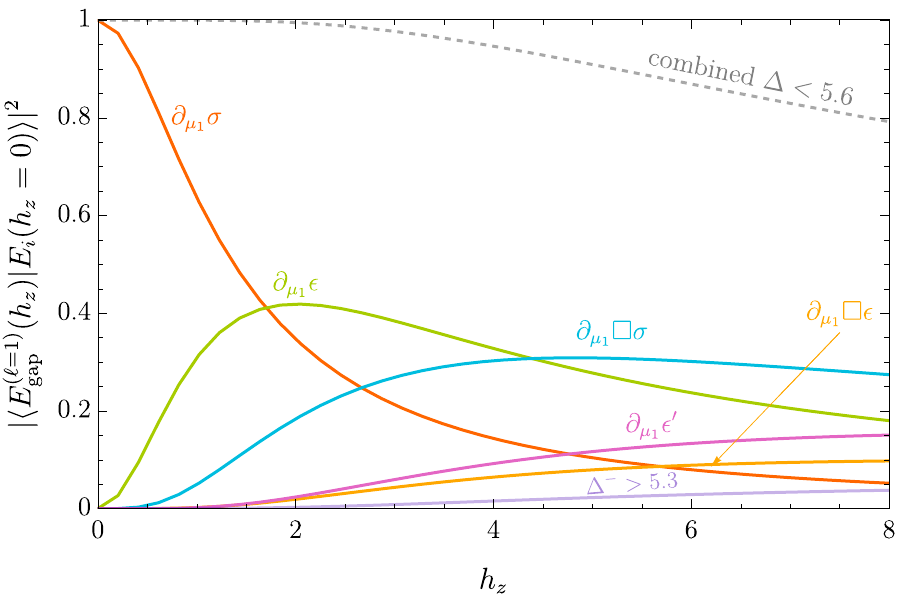}
        \caption{$\ell=1$}
    \end{subfigure}%
    \hfill
    \begin{subfigure}[t]{0.48\textwidth}
        \centering
        \includegraphics[ width=1\textwidth]{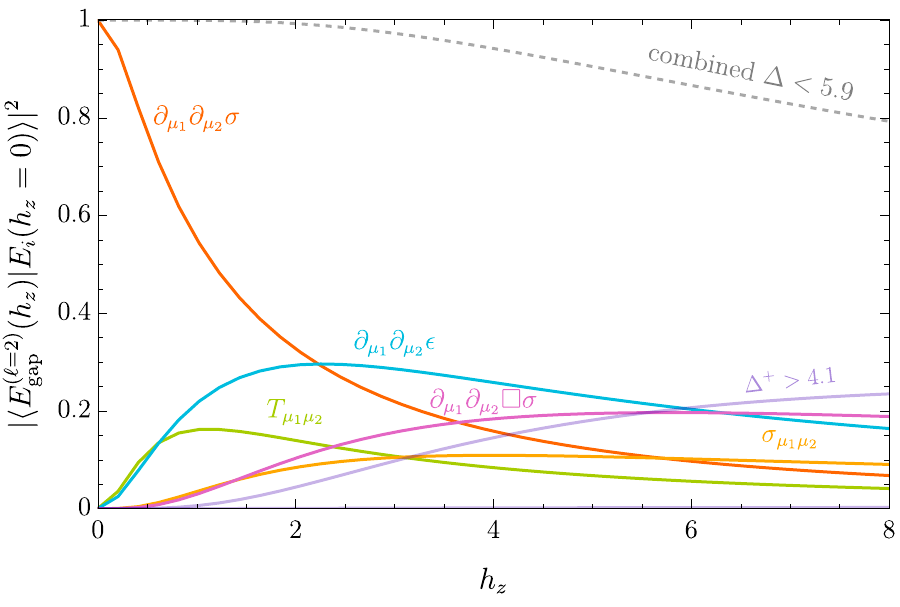}
        \caption{$\ell=2$}
    \end{subfigure}
    \\[\baselineskip]
        \begin{subfigure}[t]{0.48\textwidth}
        \centering
        \includegraphics[ width=1\textwidth]{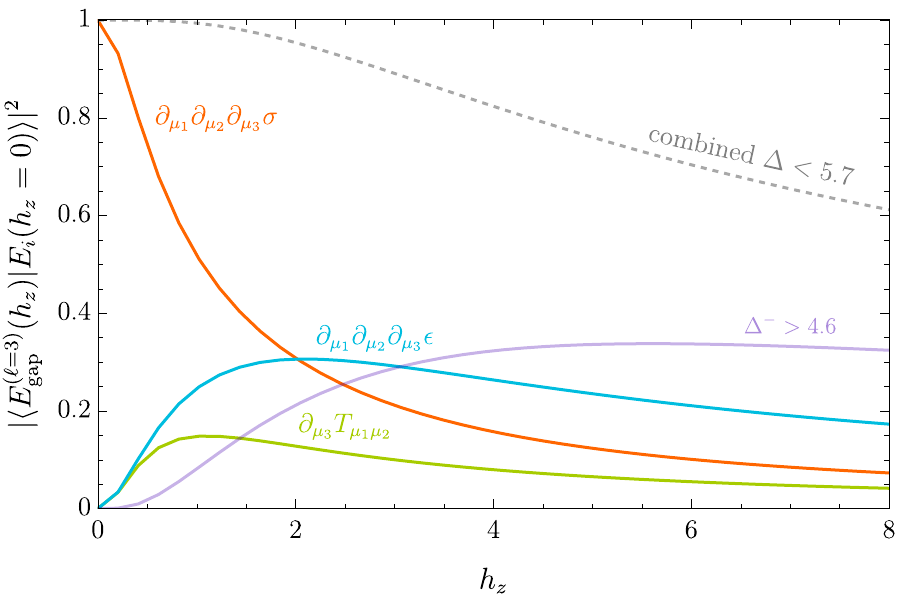}
        \caption{$\ell=3$}
    \end{subfigure}%
    \hfill
    \begin{subfigure}[t]{0.48\textwidth}
        \centering
        \includegraphics[ width=1\textwidth]{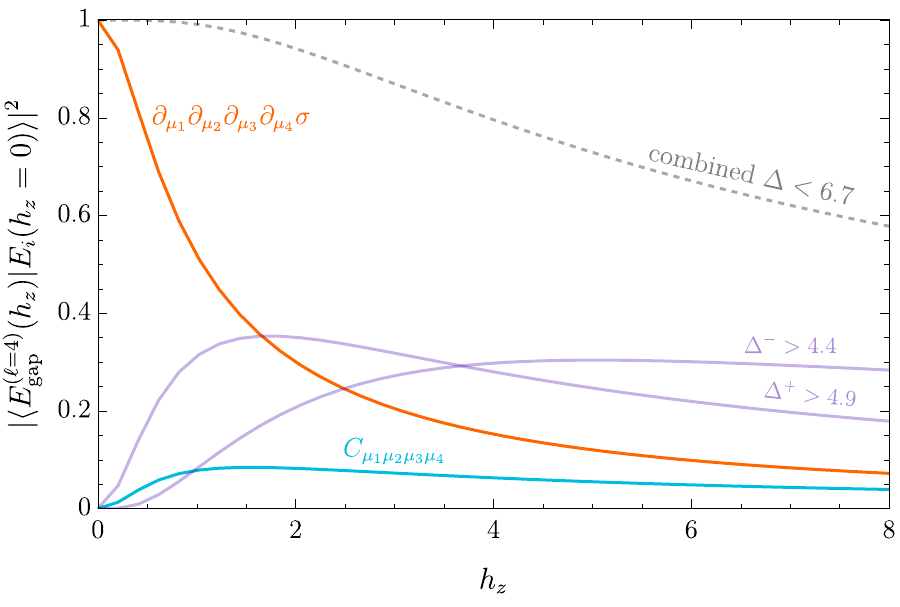}
        \caption{$\ell=4$}
    \end{subfigure}
    \caption{The CFT state content of the one-particle state in $\ell=1,2,3,4$ sectors.  Purple lines indicate $\mathbb{Z}_2$-even (+) and $\mathbb{Z}_2$-odd (-) states which have not been clearly identified (see text).  The gray dashed line shows the total squared overlap captured by the CFT states included in the analysis, obtained by summing over the spin-$\ell$ states among the 50 lowest-energy states retained a $h_z=0$. The absence of a curve associated with a given $\mathbb{Z}_2$ charge indicates that the corresponding contribution to the overlap is small. Although we cannot always disentangle the individual primary and descendant contributions, the dominant states can nevertheless be characterized. For example, in the $\ell=2$ sector, the leading $\mathbb{Z}_2$-even contributions are primarily combinations of $T'_{\mu_1\mu_2}$ and spinning descendants of $\epsilon$ and $\epsilon^\prime$.  At spin-3 the dominant contributions are $\mathbb{Z}_2$-odd and involve $\partial_{\mu_1}\partial_{\mu_2}\partial_{\mu_3}\sigma$ and $\partial_{\mu_3}\sigma_{\mu_1\mu_2}$ (by contrast, $\sigma_{\mu_1\mu_2\mu_3}$ gives only a small contribution).  In the $\ell=4$ sector, the dominant $\mathbb{Z}_2$-even contributions are combinations of the first spin-4 descendants of $\epsilon$ and $\partial_{\mu_3}\partial_{\mu_4}T_{\mu_1\mu_2}$, while the leading $\mathbb{Z}_2$-odd contributions involve $\partial_{\mu_3}\partial_{\mu_4}\sigma_{\mu_1\mu_2}$ and $\sigma_{\mu_1\cdots \mu_4}$.}
    \label{fig:overlapswithspin}
\end{figure} 
 
In preparation for the TCSA treatment, it is interesting to use the FS to explore the principle behind truncation.  Namely, that upon relevant deformation, the lowest energy particle states, are predominantly made of the lowest dimension CFT states.  However, due to the orthogonality catastrophe, as the volume is increased (or equivalently the size of the deformation in units of $R$), we also expect that one needs an increasing number of CFT states in order to properly capture the eigenstates of the IFT Hamiltonian.  It thus becomes a quantitative question, for a given $h_z$, which CFT states are needed in order to represent, for example, the one-particle state.  To answer this question we will compute the overlap between the one-particle state, computed at $N=14$ for a given value of $h_z$, and the CFT states, defined as the eigenstates at $h_z=0$.\footnote{For the scalar-sector analysis, we use ED to compute the spectrum at each value of $h_z$, keeping the lowest 150 eigenvalues. These include states with different angular momenta. For $\ell>0$, we instead restrict to the sector with $J_z=\ell$ and keep the lowest 50 eigenvalues. This difference explains why the maximum value of $\Delta$ for the summed overlaps is smaller for $\ell>0$.}

In Fig. \ref{fig:overlapspin0} we show the one-particle content in the scalar sector as a function of $h_z$. Indeed, for $h_z <4$ the one-particle state is well described by CFT states with $\Delta < 7.3$.  Note, that in this computation we have not used the special conformal generators we introduced in~\cite{Fardelli:2024qla} to construct states, and thus, some of the higher dimension states are mixtures of primaries and descendants that we do not attempt to disentangle here.  Such states, are labeled by their $\mathbb{Z}_2$ charges, but are not identified.  Despite this limitation, we can still identify the main origin of the missing overlap: the dominant contributions appear to come from linear combinations of the CFT operator $\epsilon''$ and descendants of $\epsilon$ and $\epsilon'$.

Similarly in Fig. \ref{fig:overlapswithspin} we report the overlaps for the one-particle state, but for various choices of angular momentum.  Again, states not explicitly identified are mixtures of primaries and descendants.  Interestingly, as we increase the angular momentum, the spinning one-particle state is made  mostly of states of low twist.

\section{TCSA}
\label{sec:TCSAresults}
In this section we consider an alternative approach --- the Truncated Conformal Space Approach (TCSA). This approach does not require a fuzzy sphere or any other quantum many-body model as a UV completion.  Instead, it directly builds the IFT as in~\eqref{eq:IFT-Ham}  from the 3d Ising CFT data. 
The difference between TCSA and the  FS is in how we obtain the basis and matrix elements. 
In the Fuzzy Sphere both the basis states and the matrix elements are defined in the UV theory of LLL fermions. 
In TCSA, the basis states are in one-to-one correspondence with the CFT local operators 
\eqna{
\ket{\CO} =\CO(0)\ket{\rm vac}\, ,  \qquad \bra{\CO}=\bra{\rm vac}\CO(\infty)\, .
}[]
The benefit of this basis is that each state is an eigenstate of the undeformed CFT with eigenvalue $\Delta_{\CO}$.
With a truncation on the basis which we take to be the energy cutoff $\Delta \leq \Delta_{\rm max}$, the truncated Hamiltonian is a finite dimensional matrix. 
For the case of the magnetic deformation only, we write explicitly the Hamiltonian matrix element form with respect to operators $\CO$ and $\CO'$
\eqna{
H_{\CO\CO^\prime} =\langle \CO | H_{\rm CFT} |\CO^\prime \rangle +\frac{ h_z}{V_3}  \int d\Omega\,  \langle \CO| \sigma(\Omega) |\CO^\prime \rangle\, ,
}[]
where the  first term is diagonal,  and $\sim \Delta_\CO \lsp \delta_{\CO, \CO'}$.
The matrix elements of the second term are instead related to the CFT local operator 3-point correlation functions, i.e. the OPE coefficients $\lambda_{\CO \sigma \CO'}$, with appropriate differential and integral transformations completely fixed by conformal algebra. 
Since the middle operator is always a scalar, the most generic form of primary 3-point function is known from \cite{Costa:2011mg} to take the form
\begin{equation}\label{embed-schematic}
  \langle \CO_1(P_1, Z_1) \CO_2(P_2) \CO_3(P_3, Z_3)\rangle = \sum_{a=0}^{{\rm min}(\ell_1, \ell_3)} \lambda_{\CO_1 \CO_2 \CO_3}^{(a)} \langle \CO_1(P_1, Z_1) \CO_2(P_2) \CO_3(P_3, Z_3)\rangle^{(a)}~.
\end{equation}
Thus, the Hamiltonian relies only on the CFT data $\Delta_{\CO}$ and $\lambda_{\CO\sigma\CO'}$ of the primary operators. In principle, the data is intrinsic to a particular CFT and is independent of how it is computed. In practice, since 3d Ising CFT is not integrable, we obtain the CFT data from numerical procedures such as conformal bootstrap \cite{Simmons-Duffin:2016wlq,Chang:2024whx,Reehorst:2021hmp} and the fuzzy sphere \cite{Fardelli:2026zas}. The detailed form of (\ref{embed-schematic}) is included in Appendix \ref{sec:construction_of_the_hamiltonian_in_tcsa} in embedding space coordinates $P_i$ and auxiliary
polarization vectors $Z_i$ in index-free notation of the spin indices. 
As an explicit construction example, consider the rest frame $\ell=0$. 
A state corresponds to either a scalar primary operator or a scalar descendant of any primary operator. 
Schematically, both can simply be written as
\eqna{
\square^n \partial_{\mu_1} \cdots \partial_{\mu_{\ell}} \CO_i^{\mu_1\cdots \mu_\ell} - ({\rm trace})\, ,
}[]
with $\partial_{\mu}\equiv\frac{\partial}{\partial {x_i^\mu}}$ and $\square\equiv \partial_{\mu}\partial^\mu$. 
For the spinning frame, the conformal basis states can be constructed by properly taking the descendant states that are $SO(3)$ highest-weight states and diagonalizing the Gram matrix. 
The embedding space coordinates in (\ref{embed-schematic}) have the advantage that the conformal inversion is trivial, so descendants of both the bra and the ket are given by derivatives.
Thus a matrix element with respect to two such operators can be computed as the Taylor expansion coefficients of (\ref{embed-schematic}) in coordinate space. 
The procedure and the explicit results are presented in  Appendix \ref{sec:construction_of_the_hamiltonian_in_tcsa}.

There are pros and cons with using TCSA. On the one hand, TCSA is much more efficient than the fuzzy sphere approach, as the former only captures degrees of freedom between the CFT scale and the IR mass gap scale, whereas the latter contains degrees of freedom starting at the fuzzy sphere UV scale all the way down to the QFT scale. On the other hand, the truncation of TCSA Hamiltonian is effectively a non-local regulator, which is more complicated than the fuzzy sphere local regulator. As a consequence, TCSA usually achieves the same level of convergence with a much smaller basis, but one cannot reduce the truncation error via simple scaling analysis or model the truncation effects by a simple local term. Moreover, the Hamiltonian is dense, so sparse-local tools such as DMRG are likely not applicable, but the size of the TCSA Hamiltonian considered in this work is so small that diagonalization time is negligible.

\paragraph{Results}

\begin{figure}[htbp]
\centering
\begin{minipage}{0.6\textwidth}\centering
\includegraphics[height=1\linewidth]{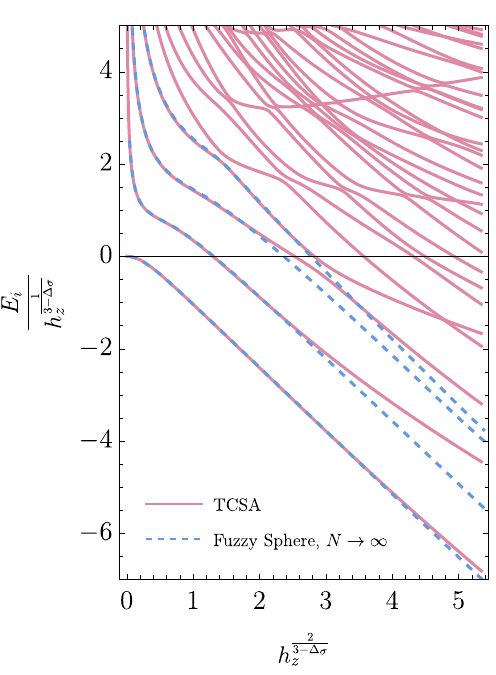} 
\end{minipage}
 \hfill
\begin{minipage}{0.39\textwidth}
\caption{\label{fig:TCSA-full}Energy eigenvalues of the TCSA Hamiltonian at our largest truncation $\Dmax^*=7.53$ (see text) in the range of coupling $0 < h_z \leq 8.0$. The TCSA eigenvalues are shown in pink solid lines, and the lowest 4 eigenvalues from Fuzzy Sphere extrapolated to $N\rightarrow \infty$ are shown in blue dashed lines in comparison. The axes are rescaled by suitable powers of $h_z$ such that the horizontal axis shows the dimensionless volume $\propto h_z^{\frac{2}{3-\Delta_\sigma}}$, and the vertical axis shows the energy in the units of the inverse of the dimensionless radius.  }
\end{minipage}
\end{figure}

We begin with an overall depiction of the TCSA results in Fig.~\ref{fig:TCSA-full}. The 3d IFT TCSA behavior is qualitatively similar to that of TCSA for 2d IFT \cite{Yurov:1991my}, as well as our own Fuzzy Sphere results in Fig.~\ref{fig:RawSpectrum}. We obtain a volume-law scaling ground state energy and a few sparse low eigenvalues. According to the previous analysis, the 3 lowest excited states are the one-particle state, the bound-state, and the lowest two-particle state. Beyond these states, the spectrum becomes dense and will eventually form a quasi-continuum at large truncation. The fact that the ground state energy obeys the volume law, which is a non-trivial power $h_z$, for a large range of couplings, requires the cooperation of many states, even within our small basis of 48 states. For the excited states, TCSA and Fuzzy Sphere results are close for small coupling. For example, both TCSA and Fuzzy Sphere agree on the kink of the 3rd excited state energy, due to avoided level collision at finite volume. 

\begin{figure}\centering
  \begin{minipage}{0.48\textwidth}\centering
  \includegraphics[width=1\textwidth]{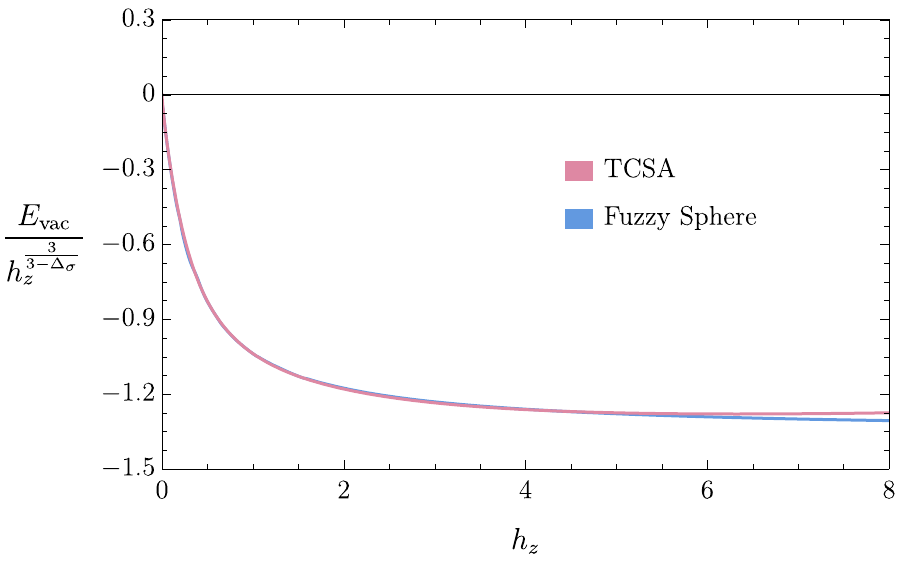}
  \end{minipage}
  \hfill
  \begin{minipage}{0.48\textwidth}\centering
  \includegraphics[width=1\textwidth]{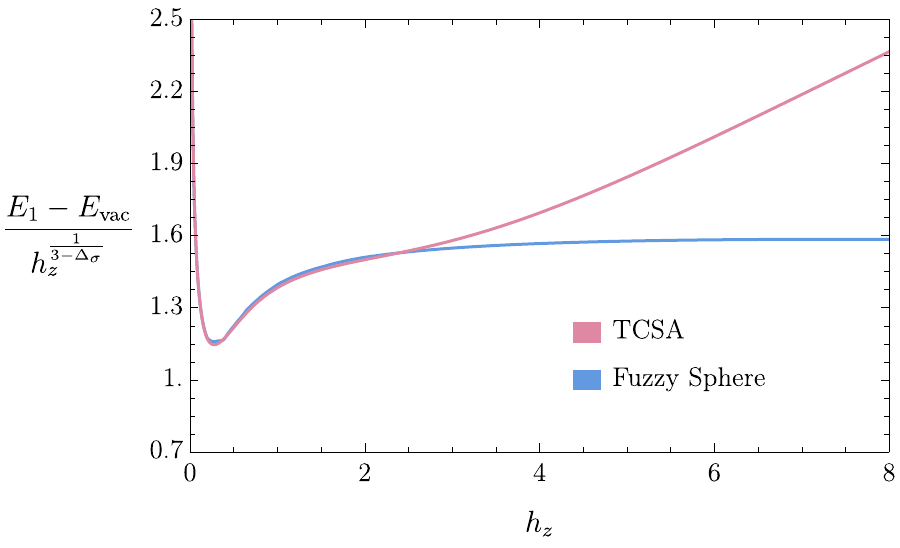}
  \end{minipage}
  \caption{\label{fig:TCSA-FS} Comparison of TCSA vs the Fuzzy Sphere for the vacuum energy (\textit{left}) and the mass gap (\textit{right}). Like Figure \ref{fig:TCSA-full}, the Fuzzy Sphere results extrapolated to $N\rightarrow \infty$ are shown with blue lines, and the TCSA results at primary truncation $\Delta_{\rm max}^* = 7.53$ are shown with pink lines.}
\end{figure} 

TCSA and Fuzzy Sphere results separate at large couplings due to different UV effects. For the Fuzzy Sphere, the UV effects have been reduced to the best of our ability via $1/N$ fitting and extrapolation, thanks to the fact that the Fuzzy Sphere regulator is local. On the TCSA side, removing truncation effects is not easy as the energy or $\Delta_{\rm max}$ truncation is non-local, and cannot be modeled with a simple $1/\Delta_{\rm max}$ fitting~\cite{Hogervorst:2014rta,Elias-Miro:2017xxf,Elias-Miro:2017tup}. Thus we keep $\Dmax$ fixed in Fig.~\ref{fig:TCSA-full}. The cutoff includes all primaries below 7.53, and their various descendants (reaching $\Delta_{\rm max} = 11.71$ in the spin-0 sector).  We refer to this set of states as $\Delta_{\rm max}^* < 7.53$, even though we keep their descendants up to $\Delta_{\rm max} = 11.71$, because as we discuss below, it is the primaries which have the most important contributions.

\begin{figure}\centering
  \begin{minipage}{0.48\textwidth}\centering
  \includegraphics[width=1\textwidth]{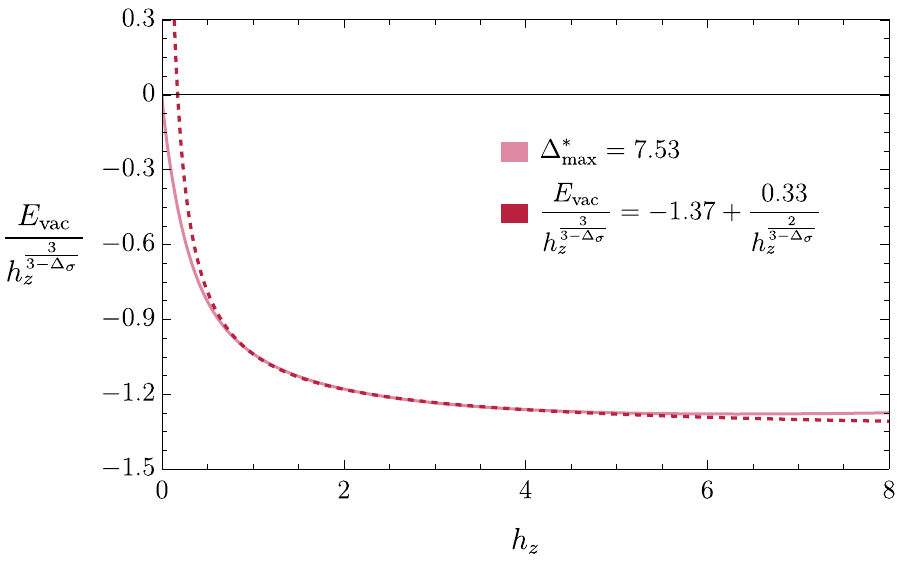}
  \end{minipage}
  \hfill
  \begin{minipage}{0.48\textwidth}\centering
  \includegraphics[width=1\textwidth]{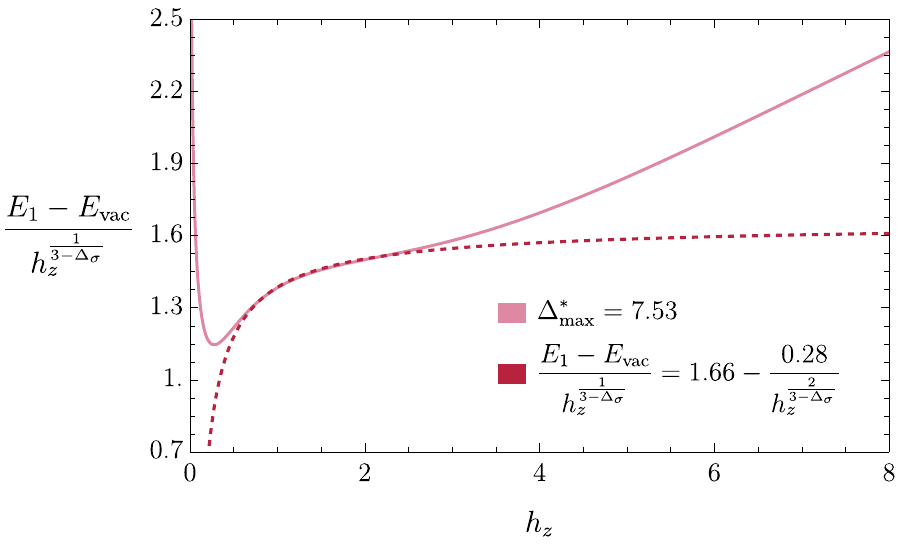}
  \end{minipage}
\caption{\label{fig:TCSA-vac-and-gap}Results from TCSA (solid line) and fit to curvature corrections (dashed line) for the vacuum energy (\textit{left}) and the energy gap (\textit{right}). }
\end{figure}

In Fig.~\ref{fig:TCSA-FS} we display the vacuum and mass-gap and focus on the comparison between TCSA and the Fuzzy Sphere. The Fuzzy Sphere has better scaling behavior at larger volume. Remarkably, the vacuum energy agrees between TCSA and Fuzzy Sphere for a wide range of couplings. For the gap, TCSA deviates from the Fuzzy Sphere properly scaling behavior for $h_z>2.7$. In addition, the EFT analysis in section \ref{sec:Hierarchy} suggests that $h_z$ cannot be too small in order to keep curvature corrections under control. Hence, we fit in the region $0.7<h_z<2.7$, the vacuum energy and gap, including the leading curvature correction in the fit.  The results are shown in Fig.~\ref{fig:TCSA-vac-and-gap}. They are in good agreement with the Fuzzy Sphere, despite a smaller range of couplings.
 
\begin{figure}\centering
  \begin{minipage}{0.48\textwidth}\centering
  \includegraphics[width=1\textwidth]{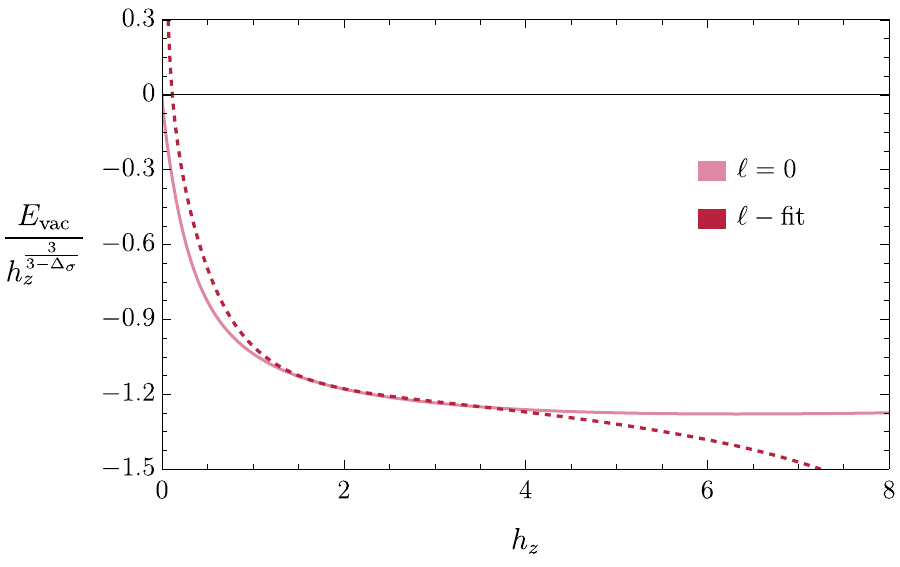}
  \end{minipage}
  \hfill
  \begin{minipage}{0.48\textwidth}\centering
  \includegraphics[width=1\textwidth]{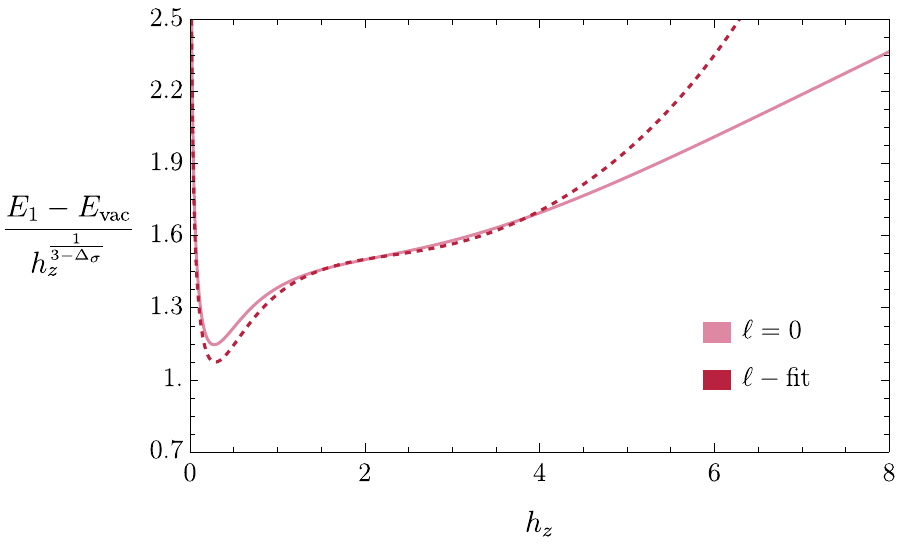}
  \end{minipage}
  \caption{Comparison of the results for the vacuum and the gap obtained from the scalar sector and the boosted ones.  The $\ell$-fits are done including $\ell=0,1,\cdots, 4$. The truncation in the spinning sector -- overall for both primary and descendant operators -- is $\Delta_{\rm max} = 11.5$, with the slight caveat that the parity odd primary operators are not included. }\label{fig:fitTCSAell} 
\end{figure} 

\begin{figure}
\centering
\includegraphics[width=.6\textwidth]{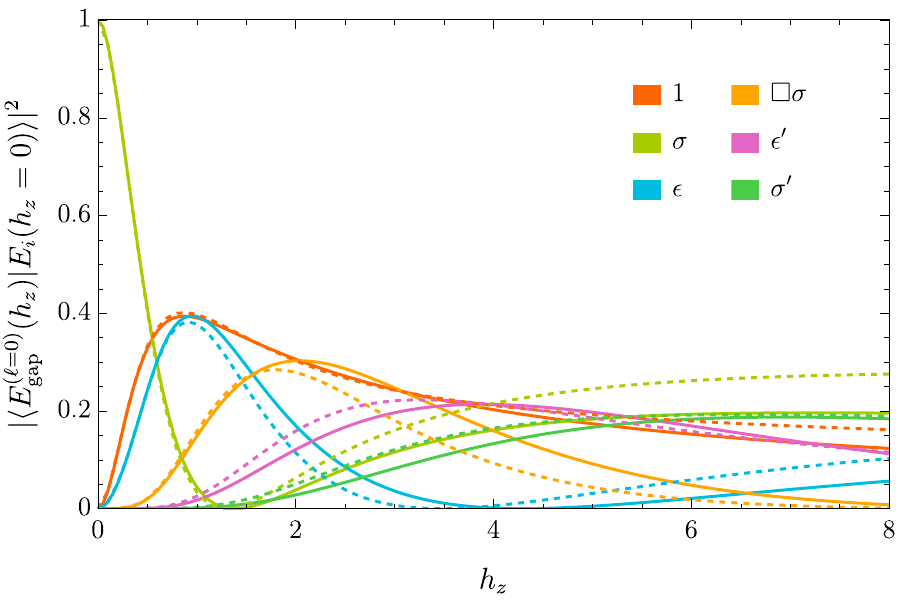}  
\caption{\label{fig:truncation-overlap}The CFT state content of the one-particle state for different values of $h_z$ in the scalar sector, with comparison between Fuzzy sphere (solid line) and TCSA (dashed line).    
}
\end{figure}

TCSA can also be used in a different angular momentum sector.  Here we use the same fitting model
\begin{align} \tag{\ref{eq:ellfit}}
 E_1(\ell)=E_{\rm vac}+\sqrt{m_{\rm gap}^2+\ell(\ell+1)+c\, \ell^2 (\ell+1)^2}\,, 
\end{align}
as we used in the Fuzzy Sphere computation. We perform the fit for each $h_z$, and extract $E_{\rm vac}$ and $m_{\rm gap}$ together from (\ref{eq:ellfit}),  and compare them  against the $\ell=0$ data in Fig.~\ref{fig:fitTCSAell}. Remarkably, the finite spin fit agrees well with $\ell=0$ data in the region where we trust the $\ell=0$ truncation. 
In Fig.~\ref{fig:truncation-overlap} we also compare the FS results for the overlap of the gap state with the CFT states at $h_z=0$ in Fig.~\ref{fig:overlapspin0} to the same quantities compute in TCSA . The overlaps are close for the two methods, with sizable discrepancies at large $h_z$, as expected.

\begin{figure}\centering
  \begin{minipage}{0.48\textwidth}\centering
  \includegraphics[width=1\textwidth]{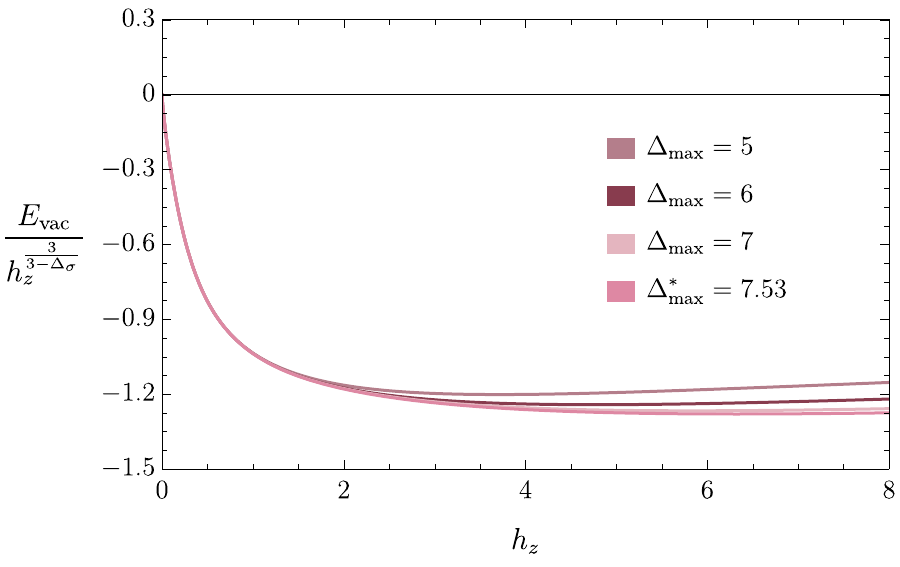}
  \end{minipage}
  \hfill
  \begin{minipage}{0.48\textwidth}\centering
  \includegraphics[width=1\textwidth]{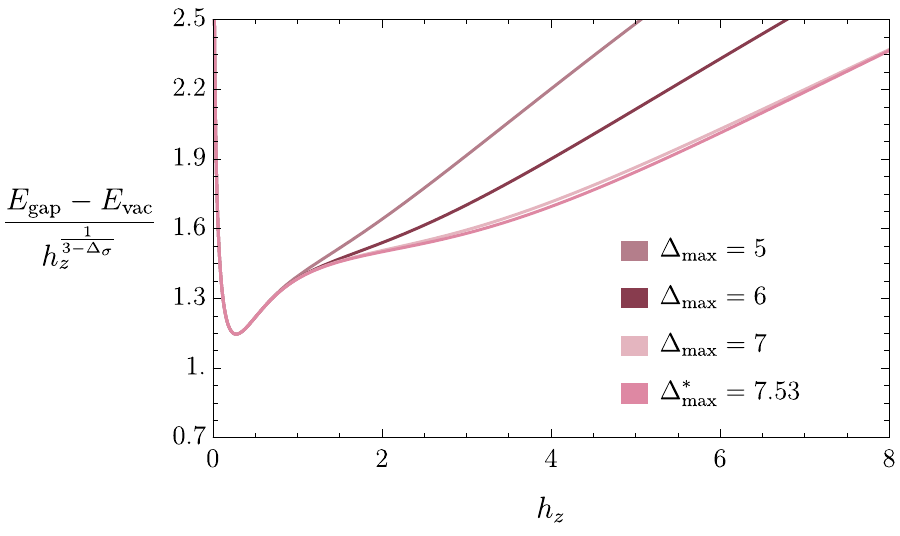}
  \end{minipage}
  \caption{\label{fig:truncation-effect}How the vacuum and the gap change as a function of $\Delta_{\rm max}$. After $\Delta_{\rm max}>7$ the changes are basically invisible.}
\end{figure} 

\paragraph{Truncation effects} 
\begin{figure}\centering
  \begin{minipage}{0.48\textwidth}\centering
  \includegraphics[width=1\textwidth]{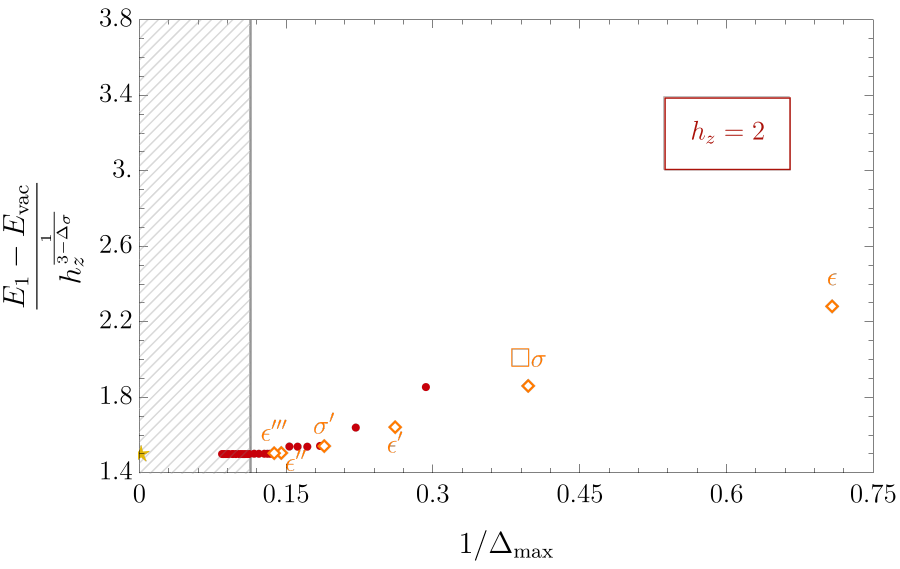}
  \end{minipage}
  \hfill
  \begin{minipage}{0.48\textwidth}\centering
  \includegraphics[width=1\textwidth]{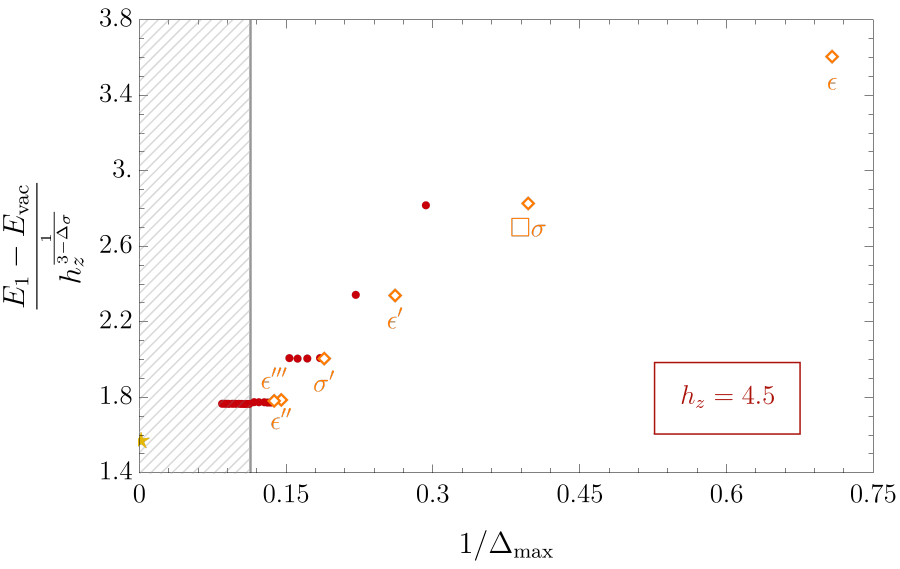}
  \end{minipage}
  \caption{\label{fig:TCSA-gap-primary-contribution}
  $(E_{1}-E_{\rm vac})$ in units of the mass gap $\mu$ as a function of $1/\Delta_{\rm max}$ at $h_z=2$ (\textit{left}) and at $h_z=4.5$ (\textit{right}).  The points are computed including all operators (descendants and primaries) appearing in~\ref{tab:dims} up to the corresponding $\Delta_{\rm max}$.  Diamonds mark the scalar primaries indicated together with $\square \sigma$.  The star represents the value we expect from the FS.  The vertical line marks the dimension of the first scalar primary of~\cite{Simmons-Duffin:2016wlq, Henriksson:2025vyi} that we do not include in the truncation, since its OPE coefficients are not currently known.
  }
\end{figure}
We also show the TCSA results at different $\Delta_{\rm max}$ truncations in Fig.~\ref{fig:truncation-effect}. 
For the vacuum energy, we see general good convergence throughout the couplings that we test, which is consistent with the agreement we see in Fig.~\ref{fig:TCSA-FS}. For the mass gap, we see something puzzling: the gap keeps improving from $\Delta_{\rm max}=3$ to $\Delta_{\rm max}=7$ but suddenly stops improving and stays essentially the same up to $\Delta_{\rm max}=11.71$. Does this mean TCSA stops converging? A detailed study of the $\Delta_{\rm max}$ convergence, shown in Fig.~\ref{fig:TCSA-gap-primary-contribution}, provides some hints. First, we see that improvement is only significant when certain type of operators pass the truncation threshold -- the primaries and $\square \sigma$ -- while most descendant operators' contribution is negligible. The $\square \sigma$ operator is almost a primary as it is related by the equation of motion $\square \phi \propto \phi^3$ in $\phi^4$ theory. Thus, the gap as a function of $\Delta_{\rm max}$ is not smooth but has stairs bordered by primaries. When we connect the edge of the stairs, the finite truncation gap appears to be smoother. The $h_z=4.5$ plot does not seem to be converging for $\Delta_{\rm max} \geq 7$ but it should be noted that we have only included primaries up to $\Delta_{\rm max}^* = 7.53$ in the scalar sector. We conjecture that the convergence should improve if we add a missing scalar primary before the truncation at $11.71$. From \cite{Simmons-Duffin:2016wlq, Henriksson:2025vyi} it seems possible that there exists a new primary at $\Delta\approx 9$, which could explain why TCSA is failing to move to the next stair.  Presumably, the issue is that although the fuzzy sphere contains an adequate approximation of this additional primary (which is the reason for its better behavior as compared to TCSA), we have yet to extract its data.\footnote{Note that we have only computed the first 150 eigenstates of the FS (in all spin sectors), and the missing scalar primary could lie beyond those.}
 
 \section{Future Directions}
 \label{sec:future}
 In this work we investigated both FS and TCSA approaches applied to the Ising Field Theory on a spatial sphere.  In particular, we tested our understanding of curvature effects, and showed that modeling these properly allows for extraction of infinite volume quantities for the first few eigenstates (given current computational reach).  There are some interesting questions regarding this specific model that merit further study.  For example, what happens in the presence of both $\epsilon$ and $\sigma$ deformations?  Also, does the fact that the bound-state is so weakly bound suggest some form of effective approximation?  Does the model also possess very narrow resonances, as it did in 2d?
 
Having demonstrated that the above technologies can be useful to study deformations of the Ising CFT in 3d, we can now attempt to apply the same techniques to deformations of other 3d CFTs.  There are many possible choices at this point~\cite{He:2026ong}, however, a class of CFTs for which studying deformations would be particularly interesting is that of gauge-theory CFTs, especially if they contain dynamical gauge bosons~\cite{Huffman:2026qqq}.  Indeed, the way FS realizes such CFTs is through ``deconfined criticality'', namely without having redundant degrees of freedom in the microscopic description.  Thus, it can be somewhat subtle to even identify the emergent CFT with a particular gauge-theory.  Deforming the CFT therefore can add much needed information.  For example, in cases when there are enough flavors, a deformation which removes the flavors at scale $\mu$, can yield gauge bosons with $g^2 \ll \mu$.  Consequently, at small $g^2R$, the theory is well described by weakly coupled gauge-bosons, which could be seen in the spectrum.
 
 Finally, one can consider deformations of CFTs which break rotational invariance, by introducing defects, or deformations localized to point on the sphere.  Such deformations have already been studied using FS in 3d, but now one can also study them using TCSA, using the same OPE data we used in this work.  It would be interesting to compare the two techniques in the future, and their respective ability to capture Boundary Conformal Theories in the IR.

\acknowledgments{We are grateful to Yin-Chen He and Slava Rychkov for helpful discussions.  Numeric results in this paper were obtained by implementing our construction within the publicly available Julia code provided at \href{https://www.fuzzified.world/}{https://www.fuzzified.world} and were performed on the Shared Computing Cluster which is administered by Boston University’s Research Computing Services.  GF,  ALF, and EK  are supported by the US Department of Energy Office of Science under Award Number DE-SC0015845, and GF was partially supported by the Simons Collaboration on the Non-perturbative Bootstrap.  This work was performed in part at Aspen Center for Physics, which is supported by National Science Foundation grant PHY-2210452. }

\appendix

\section{Construction of the Hamiltonian in TCSA}
\label{sec:construction_of_the_hamiltonian_in_tcsa}
In this appendix, we review the technical details needed to construct the Hamiltonian matrix elements in TCSA. In particular, we explain how the interaction matrix elements can be obtained from CFT three-point functions involving two external primary operators and the deformation operator. We first focus on the scalar sector, where we explain in detail how to relate three-point functions of spinning primary operators to matrix elements between their scalar descendants. We finally briefly comment on how this construction can be generalized to spinning sectors.

In radial quantization, given a CFT operator $\CO(x)$,  we define the corresponding in- and out-states schematically as
\eqna{
\ket{\CO} =\CO(0)\ket{\rm vac}\, ,  \qquad \bra{\CO}=\bra{\rm vac}\CO(\infty)\, .
}[]
The problem of determining the spectrum of the Hamiltonian in~\eqref{eq:IFT-Ham} can then be formulated as an eigenvalue problem for a finite-dimensional matrix. In TCSA, we select the space of states  (including primaries and descendants) satisfying $\Delta_{\CO}<\Delta_{\rm max}$, for which the matrix elements of the Hamiltonian are
\eqna{
H_{\CO\CO^\prime} =\langle \CO | H_{\rm CFT} |\CO^\prime \rangle + \frac{h_z}{V_3}  \int d\Omega\,  \langle \CO| \sigma(\Omega) |\CO^\prime \rangle\, .
}[]
Since the external states are CFT operators, they are by definition eigenstates of $H_{\rm CFT}$ and thus the first term is diagonal and reduces to $\Delta_{\CO} \delta_{\CO\CO^\prime}$.  The matrix elements of the deformation are instead determined by the three-point functions involving $\CO$, $\CO^\prime$, and $\sigma$.
To make this relation explicit, consider a three-point function involving two generic spinning operators, with dimensions $\Delta_1,\Delta_3$ and spins $\ell_1,\ell_3$, and a scalar operator $\CO_2$ of dimension $\Delta_2$ (which will later be identified with $\sigma$)~\cite{Costa:2011mg}:
\begin{equation}
\langle \CO_1(P_1, Z_1) \CO_2(P_2) \CO_3(P_3, Z_3)\rangle = \frac{\sum_{a=0}^{{\rm min}(\ell_1, \ell_3)} \lambda_{\CO_1 \CO_2 \CO_3}^{(a)} H_{13}^a V_{1,32}^{\ell_1 - a} V_{3,21}^{\ell_3 - a}}{(2P_1 \cdot P_2)^{\frac{\bar{h}_1 + \bar{h}_2-\bar{h}_3}{2}} (2P_1 \cdot P_3)^{\frac{\bar{h}_1 + \bar{h}_3-\bar{h}_2}{2}} (2P_2 \cdot P_3)^{\frac{\bar{h}_2 + \bar{h}_3-\bar{h}_1}{2}}}, \label{eq:TensorBasis}
\end{equation}
where $\bar{h}_i \equiv \Delta_i + \ell_i$ and we use embedding-space coordinates $P_i$ for the positions and auxiliary polarization vectors $Z_i$ to encode the spin indices.  The coefficients $\lambda_{\CO_1\CO_2\CO_3}^{(a)}$ are the independent OPE coefficients, while $H_{ij}$ and $V_{i,jk}$ are the standard embedding space tensor structures defined as 
\eqna{
H_{ij} &\equiv -2[ (Z_i \cdot Z_j)(P_i \cdot P_j) - (Z_i \cdot P_j)(Z_j \cdot P_i)]\, , \\
 V_{i,jk} & \equiv \frac{(Z_i \cdot P_j)(P_i \cdot P_k) - (Z_i\cdot P_k)(P_i \cdot P_j)}{(P_j \cdot P_k)}\, .
}[]
We can then extract the corresponding matrix elements in radial quantization by sending the operator at $P_1$ to conformal infinity and expanding the operator at $P_3$ around the origin.  A convenient way to implement the former is to first perform a conformal inversion, which in  embedding-space formalism is straightforward.  From this moment on, we therefore take $P_1$   to be the conformal inversion of the real space point $x_1$ (and similarly for  $Z_1$).   Finally,  we take the point $P_2$, where the scalar $\CO_2$ is inserted, to lie on the unit sphere, $\hat{x}_2^2=1$.  Then, the corresponding embedding space coordinates in terms of the real space coordinates are 
\begin{equation}
\begin{aligned} \label{eq:subs}
Z_1 &= (2 u_1 \cdot x_1, 0 ,u_1^A)  , \\
Z_3 &= (0 , 2 u_3 \cdot x_3, u_3^A) , \\
P_1 &= (x_1^2, 1 , x_1^A) , \\
P_3 & = (1, x_3^2, x_3^A) , \\
P_2 & = (1,1,\hat{x}_2^A) .
\end{aligned}
\end{equation}
This formalism  provides a straightforward way to extract matrix elements involving descendants of primary operators. Descendants are generated by derivatives with respect to the insertion points. Thus, a matrix element between  a level-$m_1$ descendant of $\CO_1$ and level-$m_3$ descendant of $\CO_3$ can be obtained by substituting~\eqref{eq:subs}  into the three-point function~\eqref{eq:TensorBasis}  and extracting the coefficients of $x_1^{m_1} x_3^{m_3}$ in the Taylor expansion around $x_{1,3}\sim 0$.  Finally since we are  working in index-free notation, $\CO(x,u)$,  whenever we need to  restore
indices we need to act with the Todorov operator,
\begin{align}
\CO^{\mu_1 \dots \mu_\ell}(x) &= D^{\mu_1} \dots D^{\mu_\ell}\, \CO(x, u)\, ,\\
D^A& \equiv \left(\frac{d-2}{2} + u \cdot \frac{\partial}{\partial u}\right) \frac{\partial}{\partial u_A} - \frac{1}{2}\, u^A \frac{\partial^2}{\partial u \cdot \partial u}\, ,
\end{align}
which implements the symmetric traceless projection. 
\subsection{Scalar sector}
Specializing now to the construction of the Hamiltonian matrix in the scalar sector,  given a spinning operator $\CO_i^{\mu_1\cdots\mu_\ell}$, a generic scalar descendants takes the schematic form
\eqna{
\square^n \partial_{\mu_1} \cdots \partial_{\mu_{\ell}} \CO_i^{\mu_1\cdots \mu_\ell}\, ,
}[]
with $\partial_{\mu}\equiv\frac{\partial}{\partial{x_i}^\mu}$ and $\square\equiv \partial_{\mu}\partial^\mu$.  
For scalar descendants at levels $m_i=\ell_i+2n_i$, the corresponding matrix elements can therefore be obtained directly from the three-point function as
\eqna{
&\langle  \square^{n_1} \partial_{\mu_1} \cdots \partial_{\mu_{\ell_1}} \CO_1^{\mu_1\cdots \mu_{\ell_1}}| O_2(\hat{x}_2) | \square^{n_3} \partial_{\mu_1} \cdots \partial_{\mu_{\ell_3}} \CO_3^{\mu_3\cdots \mu_{\ell_3}}\rangle=\\
& \frac{ \square_1^{n_1} \square_3^{n_3} (\partial_1 \cdot D_1)^{\ell_1}(\partial_3 \cdot D_3)^{\ell_3} \langle \CO_1 (x_1, u_1) \CO_2(x_2) \CO_3(x_3, u_3) \rangle\Big|_{x_1^{\ell_1+2n_1} x_3^{\ell_3+2n_3}}}{\lVert \square^{n_1} \partial \cdot \CO_1 \rVert \lVert \square^{n_3} \partial \cdot \CO_3 \rVert}\, ,
}[]
where the norm of a scalar descendant of a spin-$\ell$ primary  can be obtained analytically
\eqna{
\lVert \square^{n} \partial \cdot \CO \rVert ^2=\frac{ n!	  (\ell!)^2  (2 \ell)! (\bar{h})_n (\bar{h}-2 \ell
   -1)_{\ell } \Gamma (2 (n+\ell +1)) (2 \bar{h}-2 \ell -1)_{2 n}}{8^{\ell }\Gamma (n+\ell +1) (\bar{h}-\ell )_n}\, ,
}[]
with $(x)_n\equiv \frac{\Gamma(x+n)}{\Gamma(x)}$ the Pochhammer symbol.  The tables
in section~\ref{subsec:OPE data} collect the dimensions and OPE coefficients needed to
build the scalar-sector Hamiltonian used in the main text.
\subsection{Spinning sector}
The spinning sectors are more involved, the essential complication being that spinning descendants of spinning primaries are generically degenerate.\footnote{Descendants of scalar primaries are never degenerate. 
Conserved currents, i.e.  operators at the unitarity bound such as the stress tensor,  instead can either have just one descendant or none at a given spin and level. }

Let us start from a scalar primary and consider the spin-$\tilde\ell$ sector. In this case, the only
spin-$\tilde\ell$ descendants are obtained by acting $\tilde\ell$ times with
derivatives. For concreteness, let us also contracts indices of in-states with the null polarization
$u^{+A}=\tfrac{1}{\sqrt2}(1,i,0)$ (and alternatively with $u^{-A}=\tfrac{1}{\sqrt2}(1,-i,0)$ for
out-states, such that $u^+\cdot u^-=1$).  Once we have saturated all the spin indices, descendants at higher levels can be obtained from the state above by acting with powers of $\square$.  Hence
\eqna{
&\langle  \square^{n_1} \partial_{-}^{\tilde{\ell}} \CO_1| O_2(\hat{x}_2) | \square^{n_3} \partial_{+}^{\tilde{\ell}} \CO_3 \rangle=\\
& \qquad \frac{\square_1^{n_1} \square_3^{n_3} (u^-\cdot \partial_1)^{\tilde{\ell}}(u^+\cdot \partial_3)^{\tilde{\ell}}\langle \CO_1 (x_1) \CO_2(x_2) \CO_3(x_3) \rangle\Big|_{x_1^{\tilde{\ell}+2n_1} x_3^{\tilde{\ell}+2n_3}}}{\lVert \square^{n_1} \partial_-^{\tilde{\ell}} \CO_1 \rVert \lVert \square^{n_3} \partial_+^{\tilde{\ell}} \CO_3 \rVert}\, .
}[]
This quantity needs then to be integrated over the sphere, which can be conveniently done using  $u^{\pm} \cdot \hat{x}_2=\frac{e^{\pm i \phi}\sin \theta}{\sqrt{2}}$.

Now, let us consider the case where the external state is a spinning primary $\ket{\CO_{\mu_1\cdots\mu_\ell}}$
(excluding conserved currents for simplicity). Then to obtain spinning descendant we may:
\begin{itemize}
\itemsep 0em
\item act with $\square$: this operation raises the level by $+2$, but leaves the spin unchanged;
\item act with $\partial_+\equiv u^+\!\cdot\partial$: level $+1$, spin $+1$;
\item contract one index with $\partial_\mu$, i.e. $\partial\cdot\CO\equiv
(\partial\cdot D)\,\CO(x,u)$: level $+1$, spin $-1$.
\end{itemize}
All remaining indices are contracted with $u^+$, such that  $\CO_+\equiv (u^+ \cdot D)\CO(x, u)$.  
A complete, non-redundant basis at
level $m$ and spin $\tilde\ell$ is labelled by $\alpha=(n,i)$,
\begin{equation}
\ket{\alpha}\;=\;\square^{\,n}\,\partial_+^{\,i}\,(\partial\cdot D)^{\,j}\;
\CO_{\underbrace{\scriptstyle +\cdots+}_{\ell-j}}\, ,
\qquad j=\ell-\tilde\ell+i\, ,
\label{eq:basis}
\end{equation}
subject to $n\ge0$ and $0\le j\le \ell$, with level
$m=2n+2i+(\ell-\tilde\ell)$. The last relation implies the parity condition
\begin{equation}
m\;\equiv\;\ell+\tilde\ell \pmod 2\, ,
\end{equation}
and counting the allowed values of $i$ gives the degeneracy
\begin{equation}
  \CN(\ell;m,\tilde\ell)\;=\;
  1+\min\!\left(\ell,\;\tilde\ell,\;\frac{m-|\ell-\tilde\ell|}{2}\right),
  \qquad m\ \ge\ |\ell-\tilde\ell|\, ,
  \label{eq:N}
\end{equation}
and $\CN=0$ otherwise.  
As an illustration, take a spin-2 primary and $\tilde\ell=2$:
\begin{itemize}
\itemsep 0em
\item $m=0$: one state, $\ket{\CO_{++}}$;
\item $m=2$: two states, $\ket{\square \CO_{++}}$ and $\ket{\partial_+ \partial^\mu \CO_{+\mu}}$;
\item $m=4$: three states, $\ket{\square^2 \CO_{++}}$, $\ket{\square \partial_+ \partial^\mu \CO_{+\mu}}$ and $\ket{\partial_+^2 \partial^{\mu}\partial^{\nu}\CO_{\mu\nu}}$;
\item for even $m\ge4$ the count saturates at three, since further levels are
reached by acting with $\square$ on the states already enumerated at $m=4$.
\end{itemize}

Because the states~\eqref{eq:basis} are neither normalized nor mutually
orthogonal, we must build the $\CN\times \CN$ Gram matrix
$G_{\alpha\beta}=\langle\alpha|\beta\rangle$ for the in- and out-multiplets
separately. Each $G$ is obtained from the same machinery with $\CO_2$ replaced
by the identity. The interaction Hamiltonian in the
spin-$\tilde\ell$ sector between level-$m_1$ and level-$m_3$ descendants then reads
\eqna{
&\langle \ell_1; m_1, \tilde{\ell}|\CO_2(\hat{x}_2) \ell_3; m_3, \tilde{\ell}\rangle= \\
& G_1^{-\frac{1}{2}} \left( \square^{n_1}\square^{n_3}\partial_-^{i_1}\partial_+^{i_3} (D\cdot \partial_1)^{j_1}(D\cdot \partial_3)^{j_3}\langle \CO_1(x_1, u_1)\CO_2(x_2)\CO_3(x_3, u_3)\Big|_{x_1^{m_1}x_3^{m_3}}\right)G_3^{-\frac{1}{2}}\,.
}[]

So far we have implicitly assumed to work with primary operators (and descendants) which are even under  spacetime parity transformation.  \\
In the scalar sector, parity-odd and parity-even primaries never mix since one cannot build a parity-even
scalar descendant starting from a parity-odd primary. 
This is no longer true once we consider spinning descendants. In particular starting from a parity odd primary, one can generate a parity even descendant using
\begin{equation}
\tilde\partial_\mu \equiv\epsilon_{\mu\nu\rho}\,u^{+\nu}\partial^{\rho}\, ,
\end{equation}
contracted with one index of $\CO$. This raises the level by one, leaves the spin
unchanged, and flips the spacetime parity. Since two $\epsilon$'s reduce to
metrics it can be used at most once, so the basis~\eqref{eq:basis} is extended
by a single label $s\in\{0,1\}$ counting its insertion, with
$s\equiv(m+\ell-\tilde\ell)\bmod 2$ fixed by the level. The parity-even states
enumerated above are those with $s=0$. 
When we allow for parity odd primaries we need to consider also   parity-odd three-point function OPE coefficients $\lambda^{-(a)}_{\CO_1 \CO_2 \CO_3}$ as\footnote{All the results for the spinning sectors in section~\ref{sec:TCSAresults} never include parity-even descendants of parity odd primaries.}
\begin{equation}
\langle \CO_1(P_1, Z_1) \CO_2(P_2) \CO_3(P_3, Z_3)\rangle = \frac{\sum_{a=1}^{{\rm min}(\ell_1, \ell_3)}  \lambda_{\CO_1 \CO_2 \CO_3}^{-(a)} \epsilon_{13,2}  H_{13}^{a-1} V_{1,32}^{\ell_1 - a} V_{3,21}^{\ell_3 - a}}{(2P_1 \cdot P_2)^{\frac{\bar{h}_1 + \bar{h}_2-\bar{h}_3}{2}} (2P_1 \cdot P_3)^{\frac{\bar{h}_1 + \bar{h}_3-\bar{h}_2}{2}} (2P_2 \cdot P_3)^{\frac{\bar{h}_2 + \bar{h}_3-\bar{h}_1}{2}}}.
\end{equation}
with, using real space coordinates, 
\eqna{
\epsilon(Z_1, Z_3, P_1, P_2, P_3) &= \det \left( \begin{array}{ccccc}  2 u_1 \cdot x_1 & 0 & x_1^2 &  1 & 1 \\
0 & 2 u_3 \cdot x_3 & 1 & x_3^2 & 1 \\
u_1 & u_3 & x_1 & x_3 & \hat{x}_2 \end{array} \right)\, .
}[]
\newpage
\subsection{OPE data} \label{subsec:OPE data}
In this appendix we collect the scaling  dimensions $\Delta_i$,  spins $\ell_i$ and parity-even OPE coefficients
$\lambda^{(a)}_{\CO \,\sigma\,\CO^\prime}$, $a=0,\dots,\min(\ell_1,\ell_3)$ of primaries used to
build the TCSA Hamiltonian in the \textit{scalar sector} in section~\ref{sec:TCSAresults}. Note that although the scalar descendants
themselves are non-degenerate, the three-point function of two spinning
primaries with $\sigma$ still carries $\min(\ell_1,\ell_2)+1$ independent
structures, so all of them are listed.  

Conformal dimensions are collected in Tab.~\ref{tab:dims}
and are taken from the numerical bootstrap results of~\cite{Simmons-Duffin:2016wlq, Reehorst:2021hmp,Chang:2024whx}.  Tab.~\ref{tab:OPEs} lists the OPE coefficients we need, taken either from the numerical bootstrap results of~\cite{Simmons-Duffin:2016wlq} or determined from the fuzzy sphere in~\cite{Fardelli:2026zas}.  In the latter case the quoted numbers are the leading large-$N$ values, obtained by extrapolating the finite-$N$ matrix elements as described there. With respect to~\cite{Simmons-Duffin:2016wlq} we rescale the OPE coefficients of spinning operators by a factor $2^{\frac{\ell}{2}}$ in order to match our conventions for the three-point structures.
\renewcommand{\arraystretch}{1.15}
 \begin{table}[htbp] \centering
 \begin{tabular}{l l l l }
 \hline
 Operator & $\Delta$ & $\ell$ & $\mathbb{Z}_2$\\
 \hline
$ \mathds{1}$ & 0 & 0 & $+$ \\
 $\sigma$  & 0.518149 & 0 & $-$ \\
$ \epsilon$  & 1.41263 & 0 & $+$ \\
$ \epsilon^\prime$ & 3.83 & 0 & $+$ \\
 {$\sigma_{\mu_1\mu_2} $} & 4.1803 & 2 &  $-$ \\
 {$\sigma_{\mu_1\mu_2\mu_3} $} & 4.638 & 3 &  $-$ \\
$C_{\mu_1\mu_2\mu_3\mu_4}$ & 5.023 & 4 & $+$ \\
 {$\sigma^\prime $} & 5.29 & 0 &  $-$ \\
$T_{\mu_1\mu_2}^\prime$ & 5.509 & 2 & $+$ \\
 {$\sigma_{\mu_1\mu_2\mu_3\mu_4} $} & 6.1127 & 4 &  $-$ \\
$C^\prime_{\mu_1\mu_2\mu_3\mu_4}$ & 6.421 & 4 & $+$ \\
 {$\sigma_{\mu_1\mu_2\mu_3\mu_4\mu_5} $} & 6.7098 & 5 &  $-$ \\
$ \epsilon^{\prime\prime}$ & 6.89 & 0 & $+$ \\
 {$\sigma_{\mu_1\mu_2}^\prime $} & 6.99 & 2 &  $-$ \\
$T_{\mu_1\mu_2}^{\prime\prime}$& 7.08 & 2 & $+$ \\
$ \epsilon^{\prime\prime\prime}$ & 7.253 & 0 & $+$ \\
$C^{\prime\prime}_{\mu_1\mu_2\mu_3\mu_4}$& 7.386 & 4 & $+$ \\
 {$\sigma'_{\mu_1\mu_2\mu_3} $}  & 7.53 & 3 &  $-$ \\
 \hline
 \end{tabular} \caption{Conformal dimensions $\Delta$, spins $\ell$ and $\mathbb{Z}_2$ parity for the spacetime parity-even primaries we use to build the TCSA Hamiltonian in the scalar sector.  The data are taken from the numerical bootstrap results in~\cite{Simmons-Duffin:2016wlq, Reehorst:2021hmp,Chang:2024whx}.}\label{tab:dims}
 \end{table}

\begin{table}
\begin{tabular}{c|cccc}
$\boldsymbol{\sigma}$ & $\sigma$&$\sigma^\prime$ & $\sigma_{\mu_1\mu_2}$ & $\sigma_{\mu_1\mu_2}^\prime$\\
\hline
$ \mathds{1}$  &1 & 0 & 0 & 0 \\
$\epsilon$ &{\bf 1.05185} & {\bf  $-$0.057235} &{\bf  0.778319} & {\bf 0.0348} \\
$\epsilon^\prime$ & $\boldsymbol{-}${\bf 0.053012} & 1.270 & 0.397 & $-$0.823 \\
 $\epsilon^{\prime\prime}$ & {\bf $-$0.00073} & 1.350 & $-$0.118 & 0.411 \\
 $\epsilon^{\prime\prime\prime}$ & {\bf $-$0.0002} & 0.180 & 0.434 & 0.027 \\
 $T_{\mu_1\mu_2}^{\prime}$ &{\bf  0.02114} & 0.392 & \{$-$0.114,0.455,1.070\} & \{$-$0.229,0.344,1.300\} \\
  $T_{\mu_1\mu_2}^{\prime\prime}$ &
 {\bf 0.00096 }& $-$0.012 & \{0.174,1.640,0.517\} & \{0.015,$-$0.404,$-$0.223\} \\
 $C_{\mu_1\mu_2\mu_3\mu_4}$  &
 {\bf 0.276304} & 0.029 & \{$-$0.023,0.209,$-$0.224\} & \{$-$0.017,0.060,0.025\} \\
 $C_{\mu_1\mu_2\mu_3\mu_4}^{\prime}$ &
{\bf  0.00784} & 0.082 & \{0.058,$-$0.147,$-$0.809\} & \{$-$0.044,0.155,0.102\} \\
 $C_{\mu_1\mu_2\mu_3\mu_4}^{\prime\prime}$ &
 {\bf 0.009508} & 0.197 & \{$-$0.092,0.210,0.873\} & \{$-$0.231,0.284,$-$0.221\} \\
\end{tabular}
\par\bigskip
\begin{tabular}{c|cc}
$\boldsymbol{\sigma}$  & $\sigma_{\mu_1\mu_2\mu_3}$ & $\sigma_{\mu_1\mu_2\mu_3}^\prime$\\
\hline
$ \mathds{1}$ & 0 & 0 \\
 $\epsilon$ &{\bf 0.391737} & $-$0.049 \\
$\epsilon^\prime$&  0.205 & 0.677 \\
 $\epsilon^{\prime\prime}$ &$-$0.043 & $-$0.247 \\
$\epsilon^{\prime\prime\prime}$ & $-$0.201 & $-$0.187 \\
 $T_{\mu_1\mu_2}^{\prime}$ &
 \{$-$0.050,0.247,0.366\} & \{0.051,$-$0.447,0.845\} \\
 $T_{\mu_1\mu_2}^{\prime\prime}$ &
 \{0.057,0.094,$-$0.344\} & \{0.011,$-$0.401,$-$0.035\} \\
  $C_{\mu_1\mu_2\mu_3\mu_4}$  &
 \{0.015,$-$0.088,0.007,$-$0.314\} & \{0.001,0.014,0.082,0.182\} \\
  $C_{\mu_1\mu_2\mu_3\mu_4}^{\prime}$ &
 \{0.022,$-$0.018,0.483,1.920\} & \{0.020,0.018,0.393,0.047\} \\
  $C_{\mu_1\mu_2\mu_3\mu_4}^{\prime\prime}$ &
 \{$-$0.037,0.221,$-$0.121,$-$0.0002\} & \{0.008,$-$0.327,$-$0.175,0.191\} \\
\end{tabular}
\par\bigskip
\begin{tabular}{c|cc}
$\boldsymbol{\sigma}$ &  $\sigma_{\mu_1\mu_2\mu_3\mu_4}$ & $\sigma_{\mu_1\mu_2\mu_3\mu_4\mu_5}$\\
\hline
$ \mathds{1}$ & 0 & 0 \\
 $\epsilon$ &{\bf 0.430821} &{\bf  0.23711 }\\
$\epsilon^\prime$ & 0.150 & 0.059 \\
 $\epsilon^{\prime\prime}$& 0.075 & 0.052 \\
 $\epsilon^{\prime\prime\prime}$ &$-$0.112 & $-$0.102 \\
  $T_{\mu_1\mu_2}^{\prime}$ &
 \{$-$0.110,0.256,$-$0.263\} & \{$-$0.048,0.217,$-$0.004\} \\
 $T_{\mu_1\mu_2}^{\prime\prime}$ &
 \{0.145,$-$0.109,0.098\} & \{0.066,0.041,0.197\} \\
  $C_{\mu_1\mu_2\mu_3\mu_4}$  &
 \{$-$0.011,0.236,$-$0.266,0.455,0.903\} & \{0.006,$-$0.174,0.121,$-$0.076,$-$1.220\} \\
 $C_{\mu_1\mu_2\mu_3\mu_4}^{\prime}$ & 
 \{0.033,$-$0.161,$-$0.230,$-$0.101,0.079\} & \{0.008,$-$0.104,0.202,0.113,0.343\} \\
  $C_{\mu_1\mu_2\mu_3\mu_4}^{\prime\prime}$ &
 \{$-$0.068,0.368,$-$0.334,0.411,1.120\} & \{$-$0.020,0.311,$-$0.184,0.074,0.255\} \\
\end{tabular} 
\caption{OPE coefficients $\lambda^{(a)}_{\CO \sigma \CO^\prime}$ as defined in~\eqref{eq:TensorBasis}, with $\CO$ and $\CO^\prime$  primaries from  Tab.~\ref{tab:dims}, used to construct the interaction TCSA Hamiltonian in the scalar sector.  Coefficients determined by the numerical bootstrap are shown in bold~\cite{Simmons-Duffin:2016wlq}; the remaining ones have been obtained from the fuzzy sphere in~\cite{Fardelli:2026zas}.}\label{tab:OPEs}
\end{table}

\bibliographystyle{JHEP}
\bibliography{refs}

\providecommand{\href}[2]{#2}\begingroup\raggedright\begin{thebibliography}{10}

\bibitem{El-Showk:2014dwa}
S.~El-Showk, M.~F. Paulos, D.~Poland, S.~Rychkov, D.~Simmons-Duffin and
  A.~Vichi, \emph{{Solving the 3d Ising Model with the Conformal Bootstrap II.
  c-Minimization and Precise Critical Exponents}},
  \href{http://dx.doi.org/10.1007/s10955-014-1042-7}{\emph{J. Stat. Phys.} {\bf
  157} (2014) 869}, [\href{https://arxiv.org/abs/1403.4545}{{\tt 1403.4545}}].

\bibitem{Zamolodchikov:1989hfa}
A.~B. Zamolodchikov, \emph{{Integrable field theory from conformal field
  theory}}, {\emph{Adv. Stud. Pure Math.} {\bf 19} (1989) 641--674}.

\bibitem{Yurov:1989yu}
V.~P. Yurov and A.~B. Zamolodchikov, \emph{{TRUNCATED CONFORMAL SPACE APPROACH
  TO SCALING LEE-YANG MODEL}},
  \href{http://dx.doi.org/10.1142/S0217751X9000218X}{\emph{Int. J. Mod. Phys.
  A} {\bf 5} (1990) 3221--3246}.

\bibitem{Hogervorst:2014rta}
M.~Hogervorst, S.~Rychkov and B.~C. van Rees, \emph{{Truncated conformal space
  approach in d dimensions: A cheap alternative to lattice field theory?}},
  \href{http://dx.doi.org/10.1103/PhysRevD.91.025005}{\emph{Phys. Rev. D} {\bf
  91} (2015) 025005}, [\href{https://arxiv.org/abs/1409.1581}{{\tt
  1409.1581}}].

\bibitem{He:2026ong}
Y.-C. He and W.~Zhu, \emph{{A Fuzzy Sphere Journey in Critical Phenomena}},
  \href{http://dx.doi.org/10.1146/annurev-conmatphys-031424-020256}{\emph{Ann.
  Rev. Condensed Matter Phys.} {\bf 17} (2026) 1--25},
  [\href{https://arxiv.org/abs/2607.01310}{{\tt 2607.01310}}].

\bibitem{Zhu:2022gjc}
W.~Zhu, C.~Han, E.~Huffman, J.~S. Hofmann and Y.-C. He, \emph{{Uncovering
  Conformal Symmetry in the 3D Ising Transition: State-Operator Correspondence
  from a Quantum Fuzzy Sphere Regularization}},
  \href{http://dx.doi.org/10.1103/PhysRevX.13.021009}{\emph{Phys. Rev. X} {\bf
  13} (2023) 021009}, [\href{https://arxiv.org/abs/2210.13482}{{\tt
  2210.13482}}].

\bibitem{Fardelli:2026zas}
G.~Fardelli, A.~L. Fitzpatrick and E.~Katz, \emph{{Improving 3d Ising OPE
  Coefficients with Fuzzy Sphere Conformal Generators}},
  \href{https://arxiv.org/abs/2602.04958}{{\tt 2602.04958}}.

\bibitem{Lauchli:2025fii}
A.~M. L{\"a}uchli, L.~Herviou, P.~H. Wilhelm and S.~Rychkov, \emph{{Exact
  diagonalization, matrix product states and conformal perturbation theory
  study of a 3D Ising fuzzy sphere model}},
  \href{http://dx.doi.org/10.21468/SciPostPhys.19.3.076}{\emph{SciPost Phys.}
  {\bf 19} (2025) 076}, [\href{https://arxiv.org/abs/2504.00842}{{\tt
  2504.00842}}].

\bibitem{Caselle:1999tm}
M.~Caselle, M.~Hasenbusch and P.~Provero, \emph{{Nonperturbative states in the
  3-D phi**4 theory}},
  \href{http://dx.doi.org/10.1016/S0550-3213(99)00333-8}{\emph{Nucl. Phys. B}
  {\bf 556} (1999) 575--600},
  [\href{https://arxiv.org/abs/hep-lat/9903011}{{\tt hep-lat/9903011}}].

\bibitem{Caselle:2001im}
M.~Caselle, M.~Hasenbusch, P.~Provero and K.~Zarembo, \emph{{Bound states and
  glueballs in three-dimensional Ising systems}},
  \href{http://dx.doi.org/10.1016/S0550-3213(01)00644-7}{\emph{Nucl. Phys. B}
  {\bf 623} (2002) 474--492}, [\href{https://arxiv.org/abs/hep-th/0103130}{{\tt
  hep-th/0103130}}].

\bibitem{Taylor:2026wan}
J.~Taylor, M.~Yusuf and Z.~Papi{\'c}, \emph{{Fuzzy Spectroscopy of Bound States
  in Massive Quantum Field Theories}},
  \href{https://arxiv.org/abs/2608.07655}{{\tt 2608.07655}}.

\bibitem{wu1966theory}
T.~T. Wu, \emph{Theory of toeplitz determinants and the spin correlations of
  the two-dimensional ising model. i}, {\emph{Physical Review} {\bf 149} (1966)
  380}.

\bibitem{fateev1994exact}
V.~Fateev, \emph{The exact relations between the coupling constants and the
  masses of particles for the integrable perturbed conformal field theories},
  {\emph{Physics Letters B} {\bf 324} (1994) 45--51}.

\bibitem{Yurov:1991my}
V.~P. Yurov and A.~B. Zamolodchikov, \emph{{Truncated fermionic space approach
  to the critical 2-D Ising model with magnetic field}},
  \href{http://dx.doi.org/10.1142/S0217751X91002161}{\emph{Int. J. Mod. Phys.
  A} {\bf 6} (1991) 4557--4578}.

\bibitem{Fonseca:2006au}
P.~Fonseca and A.~Zamolodchikov, \emph{{Ising spectroscopy. I. Mesons at T
  {\ensuremath{<}} T(c)}},  \href{https://arxiv.org/abs/hep-th/0612304}{{\tt
  hep-th/0612304}}.

\bibitem{Zamolodchikov:2013ama}
A.~Zamolodchikov, \emph{{Ising Spectroscopy II: Particles and poles at $T >
  T_c$}},  \href{https://arxiv.org/abs/1310.4821}{{\tt 1310.4821}}.

\bibitem{Gabai:2019ryw}
B.~Gabai and X.~Yin, \emph{{On the S-matrix of Ising field theory in two
  dimensions}}, \href{http://dx.doi.org/10.1007/JHEP10(2022)168}{\emph{JHEP}
  {\bf 10} (2022) 168}, [\href{https://arxiv.org/abs/1905.00710}{{\tt
  1905.00710}}].

\bibitem{Brezin:1972se}
E.~Brezin and D.~J. Wallace, \emph{{CRITICAL BEHAVIOR OF A CLASSICAL HEISENBERG
  FERROMAGNET WITH MANY DEGREES OF FREEDOM}},
  \href{http://dx.doi.org/10.1103/PhysRevB.7.1967}{\emph{Phys. Rev. B} {\bf 7}
  (1973) 1967}.

\bibitem{Fardelli:2024qla}
G.~Fardelli, A.~L. Fitzpatrick and E.~Katz, \emph{{Constructing the Infrared
  Conformal Generators on the Fuzzy Sphere}},
  \href{https://arxiv.org/abs/2409.02998}{{\tt 2409.02998}}.

\bibitem{Hu:2023xak}
L.~Hu, Y.-C. He and W.~Zhu, \emph{{Operator Product Expansion Coefficients of
  the 3D Ising Criticality via Quantum Fuzzy Spheres}},
  \href{http://dx.doi.org/10.1103/PhysRevLett.131.031601}{\emph{Phys. Rev.
  Lett.} {\bf 131} (2023) 031601},
  [\href{https://arxiv.org/abs/2303.08844}{{\tt 2303.08844}}].

\bibitem{Zhou:2025liv}
Z.~Zhou, \emph{{FuzzifiED : Julia Package for Numerics on the Fuzzy Sphere}},
  \href{https://arxiv.org/abs/2503.00100}{{\tt 2503.00100}}.

\bibitem{Costa:2011mg}
M.~S. Costa, J.~Penedones, D.~Poland and S.~Rychkov, \emph{{Spinning Conformal
  Correlators}}, \href{http://dx.doi.org/10.1007/JHEP11(2011)071}{\emph{JHEP}
  {\bf 11} (2011) 071}, [\href{https://arxiv.org/abs/1107.3554}{{\tt
  1107.3554}}].

\bibitem{Simmons-Duffin:2016wlq}
D.~Simmons-Duffin, \emph{{The Lightcone Bootstrap and the Spectrum of the 3d
  Ising CFT}}, \href{http://dx.doi.org/10.1007/JHEP03(2017)086}{\emph{JHEP}
  {\bf 03} (2017) 086}, [\href{https://arxiv.org/abs/1612.08471}{{\tt
  1612.08471}}].

\bibitem{Chang:2024whx}
C.-H. Chang, V.~Dommes, R.~S. Erramilli, A.~Homrich, P.~Kravchuk, A.~Liu
  et~al., \emph{{Bootstrapping the 3d Ising stress tensor}},
  \href{http://dx.doi.org/10.1007/JHEP03(2025)136}{\emph{JHEP} {\bf 03} (2025)
  136}, [\href{https://arxiv.org/abs/2411.15300}{{\tt 2411.15300}}].

\bibitem{Reehorst:2021hmp}
M.~Reehorst, \emph{{Rigorous bounds on irrelevant operators in the 3d Ising
  model CFT}}, \href{http://dx.doi.org/10.1007/JHEP09(2022)177}{\emph{JHEP}
  {\bf 09} (2022) 177}, [\href{https://arxiv.org/abs/2111.12093}{{\tt
  2111.12093}}].

\bibitem{Elias-Miro:2017xxf}
J.~Elias-Miro, S.~Rychkov and L.~G. Vitale, \emph{{High-Precision Calculations
  in Strongly Coupled Quantum Field Theory with Next-to-Leading-Order
  Renormalized Hamiltonian Truncation}},
  \href{http://dx.doi.org/10.1007/JHEP10(2017)213}{\emph{JHEP} {\bf 10} (2017)
  213}, [\href{https://arxiv.org/abs/1706.06121}{{\tt 1706.06121}}].

\bibitem{Elias-Miro:2017tup}
J.~Elias-Miro, S.~Rychkov and L.~G. Vitale, \emph{{NLO Renormalization in the
  Hamiltonian Truncation}},
  \href{http://dx.doi.org/10.1103/PhysRevD.96.065024}{\emph{Phys. Rev. D} {\bf
  96} (2017) 065024}, [\href{https://arxiv.org/abs/1706.09929}{{\tt
  1706.09929}}].

\bibitem{Henriksson:2025vyi}
J.~Henriksson, S.~R. Kousvos and J.~Roosmale~Nepveu, \emph{{EFT meets CFT:
  multiloop renormalization of higher-dimensional operators in general
  {\ensuremath{\phi}}$^{4}$ theories}},
  \href{http://dx.doi.org/10.1007/JHEP05(2026)005}{\emph{JHEP} {\bf 05} (2026)
  005}, [\href{https://arxiv.org/abs/2511.16740}{{\tt 2511.16740}}].

\bibitem{Huffman:2026qqq}
E.~Huffman, Z.~Zhou, Y.-C. He and J.~S. Hofmann, \emph{{Generalizing deconfined
  criticality to 3D N-flavor SU(2) quantum chromodynamics on the fuzzy
  sphere}}, \href{http://dx.doi.org/10.1103/f6qg-q875}{\emph{Phys. Rev. D} {\bf
  114} (2026) 034503}, [\href{https://arxiv.org/abs/2602.11255}{{\tt
  2602.11255}}].

\end{thebibliography}\endgroup

\end{document}